\documentclass[fleqn,usenatbib]{mnras}

\usepackage{newtxtext,newtxmath}

\usepackage[T1]{fontenc}

\DeclareRobustCommand{\VAN}[3]{#2}
\let\VANthebibliography\thebibliography
\def\thebibliography{\DeclareRobustCommand{\VAN}[3]{##3}\VANthebibliography}

\usepackage{graphicx}
\usepackage{amsmath}
\usepackage{amsmath}
\usepackage{fontawesome5}
\usepackage{color,soul}

\usepackage{bm}
\usepackage{multicol} 
\usepackage[export]{adjustbox}
\usepackage{kantlipsum}

\title[Molecular Gas across Cosmic Time]{The Column Densities of Molecular Gas across Cosmic Time: Bridging Observations and Simulations}

\author[R. Szakacs et al.]{Roland Szakacs,$^{1}$\thanks{E-mail: roland.szakacs@eso.org}
Céline Péroux,$^{1,2}$
Martin A. Zwaan,$^{1}$
Dylan Nelson,$^{3}$
Eva Schinnerer,$^{4}$
Natalia Lahén,$^{5}$
\newauthor Simon Weng,$^{1,6,7}$
Alejandra Y. Fresco$^{8}$
\\
$^{1}$European Southern Observatory (ESO), Karl-Schwarzschild-Str. 2, 85748 Garching bei München, Germany\\
$^{2}$Aix Marseille Univ, CNRS, CNES, LAM, Marseille, France\\
$^{3}$Universitat Heidelberg, Zentrum für Astronomie, Institut für theoretische Astrophysik, Albert-Ueberle-Str. 2, 69120 Heidelberg, Germany\\
$^{4}$Max Planck Institute for Astronomy, Königstuhl 17, 69117 Heidelberg, Germany\\
$^{5}$Max Planck Institute for Astrophysics, Karl-Schwarzschild-Str. 1, D-85740, Garching, Germany\\
$^{6}$Sydney Institute for Astronomy, School of Physics, University of Sydney, NSW 2006, Australia\\
$^{7}$ATNF, CSIRO Astronomy and Space Science, PO Box 76, Epping, NSW 1710, Australia\\
$^{8}$Max-Planck-Institut für Extraterrestrische Physik (MPE), Giessenbachstr. 1, 85748 Garching bei München, Germany\\
}

\date{Accepted XXX. Received YYY; in original form ZZZ}

\pubyear{2021}

\begin{document}
\label{firstpage}
\pagerange{\pageref{firstpage}--\pageref{lastpage}}
\maketitle

\begin{abstract}
Observations of the cosmic evolution of different gas phases across time indicate a marked increase in the molecular gas mass density towards $z\sim 2-3$. Such a transformation implies an accompanied change in the global distribution of molecular hydrogen column densities ($N_{\rm{H_2}}$). Using observations by PHANGS-ALMA/SDSS and simulations by GRIFFIN/IllustrisTNG we explore the evolution of this H$_2$ column density distribution function [$f(N_{\rm{H}_2})$]. The H$_2$ (and HI) column density maps for TNG50 and TNG100 are derived in post-processing and are made available through the IllustrisTNG online API. The shape and normalization of $f(N_{\rm{H}_2})$ of individual main-sequence star-forming galaxies are correlated with the star formation rate (SFR), stellar mass (${M_*}$), and H$_2$ mass ($M_{\rm{H}_2}$) in both observations and simulations. TNG100, combined with H$_2$ post-processing models, broadly reproduces observations, albeit with differences in slope and normalization. Also, an analytically modelled $f(N)$, based on exponential gas disks, matches well with the simulations. The GRIFFIN simulation gives first indications that the slope of $f(N_{\rm{H}_2})$ might not majorly differ when including non-equilibrium chemistry in simulations. The $f(N_{\rm{H}_2})$ by TNG100 implies that higher molecular gas column densities are reached at $z=3$ than at $z=0$. Further, denser regions contribute more to the molecular mass density at $z=3$. Finally, H$_2$ starts dominating compared to HI only at column densities above log($N_{\rm{H}_2} / \rm{cm}^{-2}) \sim 21.8-22$ at both redshifts. These results imply that neutral atomic gas is an important contributor to the overall cold gas mass found in the ISM of galaxies including at densities typical for molecular clouds at $z=0$ and $z=3$.
\end{abstract}

\begin{keywords}
galaxies: evolution - quasars: absorption lines - ISM: evolution - ISM: atoms - ISM: molecules 
\end{keywords}

\section{Introduction}

While the total amount of baryons in the Universe ($\Omega_{\rm{baryons}} = \rho_{\rm{baryons}}/\rho_{\rm{crit,0}} \sim 4$ per cent, where $\rho_{\rm{crit,0}}$ is the critical density of the Universe) is well established from measurements of anisotropies in the Cosmic Microwave Background \citep{Planck_2016} and from primordial nucleosynthesis \citep{Cooke_2018}, the contribution and evolution of different gas phases remain to be probed. Especially constraints on the evolution of the phases most closely linked to star formation, namely the neutral atomic and molecular gas phases, are limited. Nonetheless, recent observations have shown first indications of how these gas phases are evolving.

The neutral atomic gas phase shows little evolution with redshift, with its comoving baryonic mass density only slightly declining as the redshift decreases [$\rho_{\rm{neutral-gas}} \sim (1+z)^{0.57 \pm 0.04}$] \citep[e.g.][]{Wolfe_2005, Noterdaeme_2009, Crighton_2015, Jones_2018, Peroux_2020, Tacconi_2020, Walter_2020}. This can be traced by the column density distribution function [$f(N_{\rm{HI}})$] across cosmic time, which describes the number of HI systems per unit column density per unit distance interval. $f(N_{\rm{HI}})$ quantifies the distribution of HI column densities on the sky and by integrating $f(N_{\rm{HI}})$ one can compute the comoving HI mass density. While $f(N_{\rm{HI}})$ of various shapes can result in the same $\rho_{\rm{neutral-gas}}$, HI-absorption in quasar spectra and emission-line measurements have revealed that the $f(N_{\rm{HI}})$ shows little to no evolution, either in shape or in normalisation \citep[e.g.][]{Zwaan_2005, Peroux_2005, Zafar_2013, Ho_2021}.

Observations calculating the comoving molecular mass density on the other hand have indicated a more radical evolution of the gas phase crucially needed for star formation. The comoving mass density of H$_2$ rises until cosmic noon ($z \sim 2-3$) where it peaks and drops towards $z = 0$ \citep[e.g.][]{Daizhong_2019, Popping_2019, Riechers_2019, Peroux_2020, Decarli_2020, Tacconi_2020, Walter_2020}. Given this evolution of the H$_2$ comoving mass density over cosmic time, changes in the normalisation or shape of $f(N_{\rm{H}_2})$ can be expected. 

Globally, the neutral atomic gas mass density is higher than that of the molecular phase \citep[][]{Peroux_2020, Tacconi_2020, Walter_2020}, but $f(N)$ helps reveal in which type of objects the neutral and molecular gas lies. HI-absorbers can be split into different categories from the Ly$\alpha$ forest for column densities $N_{\rm{HI}} \leq 1.6 \times 10^{17} \rm{cm}^{-2}$, to Lyman-limit systems (LLSs, $1.6 \times 10^{16} \leq N_{\rm{HI}} \leq 10^{19} \rm{cm}^{-2}$), to sub-damped Ly$\alpha$ absorbers (sub-DLAs, $10^{19} \leq N_{\rm{HI}} \leq 2 \times 10^{20} \rm{cm}^{-2}$), up to Damped Ly$\alpha$ absorbers (DLAs, $N_{\rm{HI}} \geq 2 \times 10^{20} \rm{cm}^{-2}$). The association between these systems and their origin is still challenging, but various works have kinematically associated LLSs, sub-DLAs and DLAs to environments like parts of the extended rotating disks, inflows and outflows of galaxies \citep[e.g.][]{Rahmani_2018a, Rahmani_2018b, Zabl_2020, Schroetter_2019, Szakacs_2021}. HI emission-line studies on the other hand \citep[e.g.][]{Zwaan_2005, Braun_2012, French_2021} can easily associate column densities with regions of galaxies like the interstellar medium (ISM) as the galaxies are completely imaged down to a given sensitivity instead of individual pencil beams. While the gas mass densities and $f(N)$ are global properties including multiple objects, comparing $f(N_{\rm{HI}})$ and $f(N_{\rm{H}_2})$ gives an indication in which regions of galaxies (e.g. the ISM, CGM, molecular clouds) neutral atomic or molecular gas dominates on average. The typical cold gas column densities for these regions are the following: Molecular Clouds: $N \geq  \times 10^{20.8} \rm{cm}^{-2}$ \citep[e.g.][]{Spilker_2021}, ISM: $N \geq  \times 10^{19}  \rm{cm}^{-2}$, CGM:$N \geq  \times 10^{14} - 10^{19} \rm{cm}^{-2}$ \citep[e.g.][]{vandevoort_2019}. Therefore, this helps us understand if neutral atomic gas is an important mass contributor in the ISM compared to molecular gas or if it is only substantial in the halos surrounding galaxies.

Today's state-of-the-art cosmological simulations enable the study of physical processes of galaxy formation for both the dark matter and baryonic component of the Universe. The results of these simulations are compared to observables to learn how well the model fits. A limitation of these simulations is that due to their large volume, the scales at which these physical processes and observables can be resolved is limited so that sub-grid models are used. The advantage that these simulations offer is the large statistical sample, as thousands of galaxies are simulated. Recently, there have been considerable efforts in modelling the cold gas phase by post-processing these simulations \citep[e.g.][]{Lagos_2015, Diemer_2018, Popping_2019}. While properties of cold gas in these simulations show various levels of (dis)agreement with observations [e.g. a higher cosmic mass density of HI and H$_2$ in IllustrisTNG compared to observations at $z=0$ \citep[][]{Diemer_2019}, tensions concerning the the cosmic metal density evolution in neutral gas in EAGLE, IllustrisTNG and L-GALAXIES 2020 \citep[][]{Yates_2021}, the lower molecular mass as a function of stellar mass and number of H$_2$ rich galaxies in IllustrisTNG compared to the ASPECS survey \citep[][]{Popping_2019}], other observables, like the HI column density distribution function have been accurately reproduced \citep[][]{Rahmati_2013}. Therefore, further studies and comparisons of these and similar observables, like the $f(N_{\rm{H}_2})$, are needed to improve the models and to design future observations.

The goal of this study is to probe the evolution of $f(N_{\rm{H}_2})$ across cosmic time. For this we compare data from observations on one hand and isolated and cosmological (magneto-)hydrodynamical simulations on the other hand. In the past, $f(N_{\rm{H}_2})$ has been studied using CO emission lines at low-$z$ \citep{Zwaan_2006} and more recently by studying composite H$_2$ QSO absorption spectra at $z$ $\sim$ 3 \citep{Balashev_2018}. High-resolution CO emission-line observations of local galaxies by the PHANGS-ALMA survey \citep{Leroy_2021} now enable us to derive $f(N_{\rm{H}_2})$ using emission lines from galaxies on scales of giant molecular clouds (GMCs). Further, state-of-the-art hydrodynamical simulations including non-equilibrium chemical networks tracking H$_2$ on-the-fly in high-resolution dwarf galaxy simulations \citep{Hu_2014, Hu_2016, Hu_2017, Lahen_2019, Lahen_2020, Lahen_2020b, Hislop_2021} and post processing the TNG100 cosmological magnetohydrodynamical simulation \citep{Marinacci_2018, Springel_2018, Naiman_2018,Nelson_2018, Pillepich_2018} enable bridging observations and simulations. Finally, we aim to compare $f(N_{\rm{HI}})$ and $f(N_{\rm{H}_2})$ to provide indications of the regions of galaxies (e.g the ISM, CGM, molecular clouds) in which the molecular or neutral atomic gas phases dominate.

The paper is organized as follows: Section \ref{sec:f_N_prescription} describes the column density distribution function $f(N)$. Section \ref{sec:bridging_obs_sim} presents the observational setup as well as the simulations used for the analysis of $f(N_{\rm{H}_2})$. Section \ref{sec:resolution} describes the resolution dependence of $f(N_{\rm{H}_2})$. Section \ref{sec:phangs_individ} presents the $f(N_{\rm{H}_2})$ of individual galaxies in the PHANGS-ALMA survey and their correlations with integrated physical properties of the galaxy. Section \ref{sec:red_evo} presents the results of the key goal of this manuscript. We describe the redshift evolution of the $f(N_{\rm{H}_2})$ derived from both observations and simulations and study their differences and similarities across cosmic time. Further, we compare $f(N_{\rm{H}_2})$ with $f(N_{\rm{HI}})$ in order to explore at which densities neutral atomic gas dominates over molecular gas in and surrounding galaxies. In Section \ref{sec:discussion} we discuss our results from the previous sections. Finally, in Section \ref{sec:conclusions} we give a summary of the findings. Throughout this paper we adopt an $H_0 = 67.74 \textrm{km s}^{-1} \textrm{Mpc}^{-1}, \Omega_\textrm{M} = 0.3089, \textrm{and} \; \Omega_\Lambda = 0.6911$ cosmology.

\section{Quantifying the distribution of column densities observed on the sky}

\label{sec:f_N_prescription}

Column densities of different chemical species or different phases are not distributed uniformly on the sky as low-density gas is more frequent within our Universe. One way to quantify the distribution of column densities is the so-called column density distribution function $f(N_s)$. It is defined such that $f(N_s)dN_sdX$ is the number of systems with column densities between $N_s$ and $N_s$ + $dN_s$ over a distance interval $dX$, where $s$ is the species one is studying (e.g. HI or H$_2$). While in the past  $f(N_s)$ have been mostly studied using absorption systems, high-resolution data of emission lines in galaxies enable an alternative way of studying the column density distribution function. Using emission-line observations one can calculate the $f(N_s)$ as follows \citep[e.g.][]{Zwaan_2005,Zwaan_2006}:

\begin{equation}
    f(N_s) = \frac{c}{H_0} \frac{\sum_i \Phi(x_i) w(x_i) A_i(\rm{log}(\mathit{N}_s))} {N_s \; \rm{ln}(10) \; \Delta \rm{log}(\mathit{N}_s)} \; \; .
    \label{eq:f_N_H2}
\end{equation}

\noindent We bin the galaxies of our samples by their stellar mass, with a bin size of $\Delta$log($M_{*,i}$/M$_{\odot}$)=0.2. $\Phi(M_{*,i})$ is the stellar mass function with $M_{*,i}$ being the central stellar mass value of the bin $i$ the corresponding galaxy is in. $w(M_{*,i}) = \frac{1}{N_{\rm{gal,i}}}$ is a weighting function taking into account the varying number of galaxies across the range log($M_{*,i}$/M$_{\odot}$) - $\Delta$log($M_{*,i}$/M$_{\odot}$)/2 to log($M_{*,i}$) + $\Delta$log($M_{*,i}$/M$_{\odot}$)/2 by calculating the reciprocal of the number of galaxies within the stellar mass bin $i$. $A_i$(log($N_s$)) is the area function describing the area corresponding to a column density in the range log($N_s)$ to log($N_s)$ + $\Delta$ log($N_s)$ for stellar mass bin $i$ in Mpc$^2$. We use $\Delta$log($N_{\rm{H}_2}$) = 0.1 in our calculations of $f(N_{\rm{H}_2})$. Finally we convert the number of systems per Mpc to that per unit redshift using $c/H_0$. The column densities in this work are in units of H$_2$ molecules cm$^{-2}$ for molecular gas and H atoms cm$^{-2}$ for neutral atomic gas. The space densities and areas are proper.

\section{Bridging Observations and Simulations}

\label{sec:bridging_obs_sim}

In order to study the H$_2$ column density distribution [$f(N_{\rm{H}_2}$)] at $z=0$ and $z=3$ we use an approach that bridges observations and simulations. We study how the state-of-the-art simulations compare to recent observations and explore if a similar evolution of the $f(N_{\rm{H}_2}$) can be seen in both of the approaches. On the observational side we use data from the PHANGS-ALMA survey\footnote{\href{https://sites.google.com/view/phangs/home}{sites.google.com/view/phangs/home}} \citep[see Sec. \ref{sec:PHANGS-ALMA_data},][]{Leroy_2021} at $z=0$ and data from \cite{Balashev_2018} (SDSS, see Sec. \ref{sec:Balashev}) at $z=3$. On the simulation side we use TNG100 of the IllustrisTNG project\footnote{\href{https://www.tng-project.org}{tng-project.org}} \citep[see Sec. \ref{sec:TNG-100_data},][]{Marinacci_2018, Springel_2018, Naiman_2018,Nelson_2018, Pillepich_2018} at both redshifts $z=0$ and $z=3$ and a high-resolution isolated dwarf galaxy simulation including a non-equilibrium chemical network from the GRIFFIN Project\footnote{\href{https://wwwmpa.mpa-garching.mpg.de/~naab/griffin-project/}{mpa-garching.mpg.de/~naab/griffin-project/}} \citep[see Sec. \ref{sec:dwarf_sim}][]{Lahen_2019, Lahen_2020, Lahen_2020b} meant to represent a low-redshift dwarf galaxy.

\subsection{Resolved Molecular Gas in Local Galaxies}
\label{sec:PHANGS-ALMA_data}

State-of-the-art mm- and radio-telescopes like the Atacama Large Millimeter/submillimeter Array (ALMA) have enabled the astronomical community to study the coldest gas in the Universe with unprecedented spatial and spectral resolution. One of the surveys making use of these technological advances is the PHANGS-ALMA survey \citep{Leroy_2021}. This survey is the first cloud-scale ($\sim 100$ pc) survey aimed at studying the physics of molecular gas within the local galaxy population and targets galaxies that lie on or near the $z$ = 0 main sequence of star-forming galaxies with a stellar mass range of $10^{9} \; \rm{M}_{\odot} < M_* < 10^{11} \; \rm{M}_{\odot}$. PHANGS-ALMA quantifies the physics of star formation and feedback at giant molecular cloud scales and further connects them to galaxy-scale properties and processes \citep{Leroy_2021}.  Further, additional state-of-the-art multi-wavelength data are provided by the PHANGS-MUSE \citep[][]{Emsellem_2021} and PHANGS-HST surveys \citep{Lee_2021}, which will study the ionized gas, stellar populations and characterize stellar clusters of the objects observed by the PHANGS-ALMA survey.

We make use of the highly resolved CO(2--1) data of the PHANGS-ALMA Survey \citep{Leroy_2021} in order to constrain the global and local $f(N_{\rm{H}_2}$) in the range log($N_{\rm{H_2}}/\rm{cm}^{-2}) \sim 19.5$ to $24$ at $z$=0. We use a pixel-by-pixel analyzed sample consisting of 70 galaxies from \cite{Sun_2020}. The stellar mass distribution of the sample can be seen in Fig. \ref{fig:stell_distr}. In summary, the CO(2--1) data were analysed by \cite{Sun_2020} as follows: The cubes were convolved to a common spatial resolution of 150 pc and 1 kpc. Then the data cubes were masked to only include voxels that contain emission detected with high confidence. Those cubes were finally integrated to create integrated intensity maps. The integrated maps were then used to derive the molecular gas surface density for each pixel as follows:\footnote{Surface density table for 150 pc can be found at \href{https://www.canfar.net/storage/list/phangs/RELEASES/Sun_etal_2020b}{canfar.net/storage/list/phangs/RELEASES/Sun\_etal\_2020b , datafileB1}. 1 kpc table provided by authors of \cite{Pessa_2021}.}

\begin{equation}
    \Sigma_{\rm{mol}} = \alpha_{\rm{CO}} \; R^{-1}_{21} \; I_{\rm{CO}} \; \; ,
\end{equation}

\noindent here $R_{21} = 0.65$ is the CO(2--1)-to-CO(1--0) line ratio \citep{Leroy_2013, den_Brok_2021} and $\alpha_{\rm{CO}}$ is the metallicity-dependent CO-to-H$_2$ conversion factor taken as:

\begin{equation}
    \alpha_{\rm{CO}} = 4.35 \; Z'^{-1.6} \; \rm{M}_{\odot} \; \rm{pc}^{-2} \; (\rm{K} \; \rm{km} \; \rm{s}^{-1})^{-1} \; \;\ ,
\end{equation}

\noindent where $Z'$ is the local ISM metallicity in units of the solar value. The local $Z'$ is estimated using the global stellar mass, effective radius and the stellar mass metallicity relation by \cite{Sanchez_2019} combined with a metallicity gradient \citep{Sanchez_2014}. For more details see \cite{Sun_2020}. For the error calculation we additionally compute the surface density using the constant $\alpha_{\rm{CO}} = 4.3 \; \rm{M}_{\odot} \; \rm{pc}^{-2} \; (\rm{K} \; \rm{km} \; \rm{s}^{-1})^{-1}$ of the Milky Way \citep{Bolatto_2013}. Measurement uncertainties are omitted as they are negligible compared to the uncertainties of the different $\alpha_{\rm{CO}}$ conversion factors used.

With this sample we are able to constrain $f$($N_{\rm{H}_2}$) at $z=0$. We convert the derived surface densities to column densities using:

\begin{equation}
    N_{\rm{H}_2} = \frac{\Sigma_{\rm{H}_2}}{\rm{M}_{\rm{{H}_2-molecule}}} \; \; ,
\end{equation}

\noindent with $\Sigma_{\rm{H}_2}$ in units of kg / cm$^{-2}$.

We then follow equation \ref{eq:f_N_H2} to calculate $f$($N_{\rm{H}_2}$) and use two stellar mass functions by \cite{Weigel_2016} as our space density function. The first stellar mass function is that of the entire sample, and the second is one for late-type galaxies only as the PHANGS sample mostly consists of late-type galaxies on the star-forming main-sequence (see Table 5 in \cite{Weigel_2016} for the schechter parameters).

\cite{Sun_2018} estimates the 100 per cent completeness surface density limit for a sub-sample of galaxies in the PHANGS-ALMA sample to be log($\Sigma_{\rm{H}_2}/\rm{M}_{\odot} \rm{pc}^{-2}$) = 10 - 100 at 120 pc resolution. This translates to a column density completeness limit of log($N_{\rm{H}_2}/\rm{cm}^{-2}$) = 20.8 - 21.8. At 150 pc resolution the completeness limit is expected to be lower. We therefore use a conservative estimate of log($N_{\rm{H}_2}/\rm{cm}^{-2}$) = 21.6 for 100 per cent completeness of the full PHANGS-ALMA sample.

\subsection{Absorption Lines as a Probe for the H\texorpdfstring{$\bm{_2}$}{\_2} Column Density Distribution at High Redshifts}

\label{sec:Balashev}

At high redshifts it is currently challenging to observe H$_2$ directly or resolve CO emission lines in galaxies at spatial scales similar to the PHANGS-ALMA survey. Therefore one has to resort to another approach to study the H$_2$ column density distribution. H$_2$ imprints resonant electronic absorption bands in the UV and so studying absorption systems is a promising way of studying $f(N_{\rm{H}_2}$) at high redshifts. H$_2$ absorption lines are usually found within Lyman-$\alpha$ absorption systems, so called Damped-Lyman Alpha absorbers (DLAs). It is time consuming to detect these H$_2$ absorbers, due to the low detection rate of $\leq 10$ per cent. For these reasons \cite{Balashev_2018} use composite spectra of DLAs by \cite{Mas-Ribas_2017}, which are based on $\sim 27 000$ DLAs from SDSS \citep{Noterdaeme_2012} in order to detect the weak mean signature of H$_2$ at $z\sim3$. \cite{Balashev_2018} revert to these composite H$_2$ spectra in order to fit a $f$($N_{\rm{H}_2}$) in the range log($N_{\rm{H}_2} / \rm{cm}^{-2}$) = 18-22 on which in turn they fit the observed composite line profiles. 

\subsection{Cosmological Simulations providing Large Statistical Samples}

\label{sec:TNG-100_data}

Cosmological simulations provide large statistical samples for studies of galaxy evolution. One of these simulations is TNG100 of the IllustrisTNG project \citep{Marinacci_2018, Springel_2018, Naiman_2018,Nelson_2018, Pillepich_2018}. TNG100 is a state-of-the-art gravomagnetohydrodynamics (MHD) cosmological simulation including a comprehensive model for galaxy formation physics \citep{Weinberger_2017, Pillepich_2018} within a 75000 ckpc/$h$ sized box using the AREPO code \citep{Springel_2010}. IllustrisTNG aims to study the physical processes that drive galaxy formation and to study how galaxies evolve within large scale structures.

We aim to exploit the large sample size of the TNG100 simulation in order to compare the observed $f(N_{\rm{H}_2})$ at $z=0$ and $z=3$. While TNG50 offers higher resolution, we choose TNG100 as our fiducial model due to two reasons: 1) The SMF, which is an important parameter in our calculations, is closer to observations for TNG100 than for TNG50. 2) To enable future comparisons with the EAGLE cosmological simulation \citep[][]{Schaye_2015}, as TNG100 is the closest in terms of resolution to the EAGLE 100 Mpc box simulation.

The molecular gas phase in current large-scale cosmological simulations is challenging to assess. Using chemical networks to track H$_2$ on-the-fly is computationally time consuming due to the complex physics involved and the high resolution needed in order for the H$_2$ mass fraction to converge within the forming molecular clouds \citep[$\sim 0.12$ pc,][]{Seifried_2017}. In order to capture the unresolved physics, one has to revert to sub-grid models, which split the cold hydrogen component in the simulations into a neutral atomic and molecular component. We use the H$_2$ post-processing catalogs of \cite{Popping_2019} for TNG100, for which three different models are available. The used models are by \cite{Blitz_2006, Gnedin_2011, Krumholz_2013}. The model by \cite{Blitz_2006} is a pressure-based empirical fit based on a sample of 14 local spiral and dwarf galaxies that have measured atomic, molecular and stellar surface densities. Using this sample they find a nearly linear relation between the hydrostatic pressure and the ratio of molecular to atomic gas. \cite{Gnedin_2011} designed a phenomenological model for the formation of molecular hydrogen, which is dependent on the gas density, dust-to-gas ratio and the far-UV radiation flux. This model was tested on cosmological simulations by \cite{Gnedin_2009}. Finally, the model by \cite{Krumholz_2013} is a column density, metallicity and radiation field dependent relation for splitting the cold hydrogen component in simulations.

\begin{figure}
    \includegraphics[width=\linewidth]{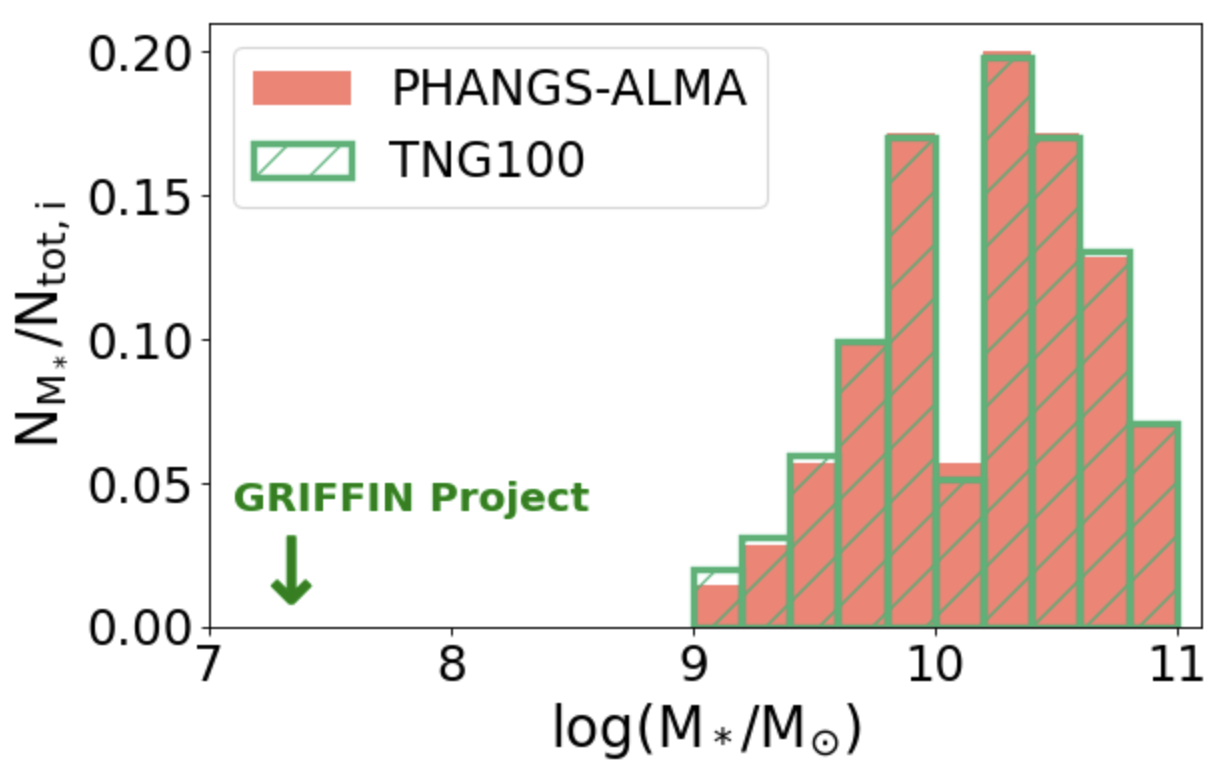}
    \caption{The stellar mass distribution of the PHANGS-ALMA survey (red) and the matching TNG100 sample (green-hatched). For each stellar mass bin we additionally only select galaxies in TNG100 with similar SFRs as found in the corresponding stellar mass bin in PHANGS-ALMA.}\label{fig:stell_distr}
  \end{figure}

In our study we select central galaxies at $z=0$ and $z=3$. We add the ability to generate post-processed H$_2$ column density ($N_{\rm{H}_2}$) maps based on \cite{Popping_2019} for TNG50 and TNG100 using the sub-grid models mentioned above, which split the cold hydrogen within those galaxies into atomic and molecular components, to the IllustrisTNG online API\footnote{\href{https://www.tng-project.org/api/}{tng-project.org/api/}}. We use this functionality to create H$_2$ column density ($N_{\rm{H}_2}$) maps at 150 pc and 1 kpc resolution. These maps are generated by projecting gas cells as adaptively sized SPH kernels. The kernel size parameter is set to $h_{\rm{kernel}} = 2.5 r_{\rm{cell}}$. With $r_{\rm{cell}}$ being the cell size determined by using the Voronoi cell volume: $r_{\rm{cell}} = (3 V_{\rm{cell}}/4\pi)^{1/3}$. We use the same projection direction for every galaxy (z-axis of the simulation) and only consider gas cells gravitationally bound to the selected subhalos within a fixed $200 \times 200$ kpc box. This method reproduces $f(N_{\rm{H}_2})$ derived from the full box of TNG100 when using the same resolution for the derivation as in \cite{Klitsch_2019}. Therefore, we do not expect this choice to affect $f(N_{\rm{H}_2})$ as compared to using a box size dependent on halo properties. At $z=0$ we selected a PHANGS-ALMA survey-like sample within TNG100. For this we select $\sim 570$ galaxies within a stellar mass of $10^9$ and $10^{11}$ M$_{\odot}$. We match the PHANGS-ALMA sample stellar mass distribution (see Fig. \ref{fig:stell_distr}). Further, for each 0.2 dex stellar mass bin we select galaxies with similar star formation rates as in the PHANGS-ALMA sample. We note that when selecting a sample of $\sim 700$ central galaxies with stellar masses between $10^9$ and $10^{12.6}$ M$_{\odot}$ and no star-formation selection criterion we derive very similar results in TNG100 as for the PHANGS-ALMA-like sample, with the only differences being the slightly higher column densities reached in the larger sample ($\sim 0.2$ dex higher) and a slightly higher normalization at column densities above log($N_{\rm{H}_2}/\rm{cm}^{-2}$) $\sim 21.5$, which is likely due to the larger size of the additional galaxies. At $z=3$ we select $\sim 550$ galaxies with stellar masses between $10^9$ and $10^{11.8}$ M$_{\odot}$. Therefore this includes galaxies between the resolution limit of TNG100 up to the highest stellar mass limit of TNG100. We do not set any constraints on the SFR of the galaxies, as the observational $f(N_{\rm{H}_2}$) is based on H$_2$ absorption line studies, where we do not have any SFR information.

Following equation \ref{eq:f_N_H2} we use the derived H$_2$ column density maps to calculate $f(N_{\rm{H}_2}$). For the space density function we use the Stellar Mass Function (SMF) of the simulation box itself. At $z=0$ the Schechter parameters are: log($M^*$/M$_{\odot}$) = 11.27, log($\Phi_1^*$/$h^3$ Mpc$^{-3}$) = -3.31, log($\Phi_2^*$/$h^3$ Mpc$^{-3}$) = -3.28, $\alpha_1$ = -1.36, $\alpha_2$ = -1.36. At $z=3$ the double Schechter parameters are: log($M^*$/M$_{\odot}$) = 10.83, log($\Phi_1^*$/$h^3$ Mpc$^{-3}$) = -3.84, log($\Phi_2^*$/$h^3$ Mpc$^{-3}$) = -3.58, $\alpha_1$ = -0.29, $\alpha_2$ = -1.64. 

\begin{figure*}
  \includegraphics[width=1.0\textwidth]{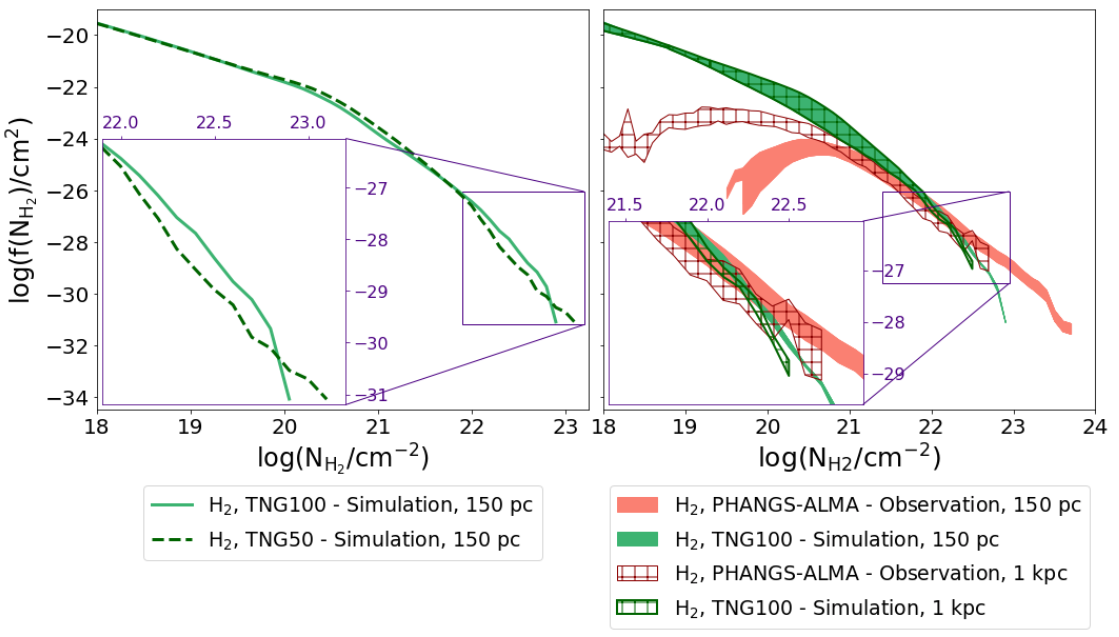}
  \caption{\textbf{Left:} $f(N_{\rm{H}_2}$) derived from TNG100 and TNG50. Higher column densities are reached in TNG50, as a higher resolution enables the simulation to reach higher gas densities. Especially at column densities above log(${N_{\rm{H}_2} / \rm{cm}^{-2}) \sim 22}$ the $f(N_{\rm{H}_2}$) differs. This indicates that H$_2$ in TNG100 is not converged for those column densities. \textbf{Right:} $f(N_{\rm{H}_2})$ derived from PHANGS-ALMA (red bands) and TNG100 (green bands) data at map resolutions of 150 pc (filled) and 1 kpc (hatched). The $f(N_{\rm{H}_2})$ both in observations and simulations show a map resolution dependence. In TNG100 this effect arises due to the averaging of column densities over a larger area. Very high column densities are usually detected on small scales (much smaller than 1 kpc), leading to a dilution of high column densities. In PHANGS-ALMA this effect is additionally combined with sensitivity and incompleteness specifics of the observations. The core of the distribution [log($N_{\rm{H}_2} / \rm{cm}^{-2}) = 21 \; \rm{to} \; 22$] is robust to resolution effects in both simulations and observations. In general, high resolution observations and simulations are needed to resolve column densities typically found in very dense environments like molecular clouds.} 
  \label{fig:f_N_150_1000pc}
\end{figure*}

\subsection{Molecular Gas in Highly Resolved Simulations of Individual Galaxies}

\label{sec:dwarf_sim}

An alternative approach to studying molecular gas in simulations is to use highly resolved simulations of individual isolated galaxies, which include non-equilibrium chemical networks that track H$_2$ on-the-fly throughout the simulation. Although this is currently mostly limited to dwarf galaxies, the advantage of these simulations is a more accurate representation of H$_2$ due to a non-equilibrium chemical network.

One of these simulations is the high-resolution isolated dwarf simulation from the GRIFFIN Project \citep{Lahen_2019, Lahen_2020, Lahen_2020b} with a stellar mass of log$(M_* / \rm{M}_{\odot}) \sim 7.3$. The simulation is based on the smoothed particle hydrodynamics tree code GADGET-3 \citep{Springel_2005} with the gas dynamics modelled using the SPH implementation SPHGal \citep{Hu_2014, Hu_2016, Hu_2017}. The simulation resolves individual massive stars at sub-parsec resolutions and includes a non-equilibrium chemical network based on \cite{Nelson_1997, Glover_2007, Glover_2012}. The chemical network follows the abundances of six chemical species for cooling processes at low temperatures ($<3 \times 10^3$ K, most importantly H$_2$). Further, the simulation includes star formation, an interstellar radiation field and stellar feedback prescriptions. A detailed discussion of the isolated dwarf simulation is given in \cite{Hu_2016, Hu_2017}.

For the calculation of $f$($N_{\rm{H}_2}$) we time- and inclination-average the isolated dwarf galaxy simulation. Therefore we produce H$_2$ column density maps with all possible lines of sight and slightly varying total H$_2$ masses using the analysis tool PYGAD \citep[][]{Roettgers_2020}. First we create H$_2$ column density maps by using snapshots over a time range of $\sim 300$ Myrs. For each of these snapshots we create H$_2$ column density maps at resolution of 150 pc with inclinations between 0 and 90$^{\circ}$ in $\Delta$cos($i$) = 0.05 steps. We then follow equation \ref{eq:f_N_H2} to calculate $f$($N_{\rm{H}_2}$) by using these H$_2$ column density maps and use the \cite{Weigel_2016} SMF of the entire sample for the normalization of $f(N_{\rm{H}_2}$) following the prescription described in Sec. \ref{sec:f_N_prescription} \citep[for the Schechter parameters see Table 5 in][]{Weigel_2016}. The $f$($N_{\rm{H}_2}$) is therefore calculated using a single stellar mass bin (as the stellar mass of the simulated dwarf galaxy does not evolve much over time). However, galaxies of this stellar mass are not represented in the PHANGS-ALMA and TNG100 sample, so we can not directly compare the $f$($N_{\rm{H}_2}$) of similar galaxies.

\section{A resolution-dependent H\texorpdfstring{$\bm{_2}$}{\_2} column density distribution function}

\label{sec:resolution}

We test how $f(N_{\rm{H}_2})$ depends on the resolution of the data used for its calculation. First we study how $f(N_{\rm{H}_2})$ depends on the resolution of the simulation by comparing $f(N_{\rm{H}_2})$ in TNG50 and TNG100 from the Illustris project. Then we compare how the resolution of the $N_{\rm{H}_2}$-maps from both observations and simulations affects $f(N_{\rm{H}_2})$.

\subsection{\texorpdfstring{$\bm{f(N_{\rm{H}_2})}$}{f(N\_H2)} - Dependence on the Resolution of Simulations}
\label{sec:TNG50_100}

Here we compare the $f(N_{\rm{H}_2}$) derived from TNG100 with TNG50 \citep[][]{Pillepich_2019, Nelson_2019} using the \cite{Gnedin_2011} H$_2$ model. TNG50 has a box length of 51.7 Mpc and $2 \times 2160^3$ resolution elements, while TNG100 has a box length of 110.7 Mpc and $2 \times 1820^3$ resolution elements. Therefore TNG50 gives us an indication how a higher resolution simulation affects $f(N_{\rm{H}_2}$). 

In Fig. \ref{fig:f_N_150_1000pc} (left) the $f(N_{\rm{H}_2}$) derived from TNG100 and TNG50 at $z=0$ using a 150 pc resolution of the post-processed column density map are displayed. TNG50 extends to higher column densities compared to TNG100. The finer resolution reaches higher gas densities and in turn higher column densities. Further, at column densities above log(${N_{\rm{H}_2} / \rm{cm}^{-2}) \sim 22}$ the $f(N_{\rm{H}_2}$) in TNG50 initially displays a steep drop with a subsequent flattening of the $f(N_{\rm{H}_2}$). These differences indicate that the H$_2$ column densities are not converged in this region. Given these differences we would expect higher resolution simulations to reach even higher column densities, and possibly also affect the shape in the region above log(${N_{\rm{H}_2} / \rm{cm}^{-2}) \sim 22}$. We note that IllustrisTNG uses the sub-grid model of \cite{Springel_2003} for the star-forming ISM. Independent of resolution, the sub-grid model begins star formation at ISM densities of 0.1 cm$^{-3}$ preventing the simulation from resolving the cold gas phase and subsequently the formation of molecular clouds. Due to this the model itself is limited by the sub-grid ISM model and a higher resolution is only sensitive up to the limitations of the model. The resolution tests, however, indicate that the sub-grid model is not the limiting factor in terms of densities reached at the resolution of TNG100 since $f(N_{\rm{H}_2}$) is not converged at high column densities. However, modifying the model to treat the multiphase ISM more realistically will likely affect the results.

\subsection{\texorpdfstring{$\bm{f(N_{\rm{H}_2})}$}{f(N\_H2)} - Dependence on the Resolution of \texorpdfstring{$\bm{N_{\rm{H}_2}}$}{N\_H2} Maps}

We compare how $f(N_{\rm{H}_2})$ depends on the resolution of observed and simulated H$_2$ column density ($N_{\rm{H}_2}$) maps. We calculate $f(N_{\rm{H}_2}$) using 150 pc and 1 kpc map resolution CO(2--1) data from the PHANGS-ALMA survey and a sample of galaxies from TNG100.

The $f(N_{\rm{H}_2})$ for these data sets and resolutions are displayed in Fig. \ref{fig:f_N_150_1000pc} (right). Differences in both shape and the column density range are found between the different map resolutions in both TNG100 and the PHANGS-ALMA survey. The $f(N_{\rm{H}_2})$ derived from PHANGS-ALMA shows a more substantial map resolution dependence compared to TNG100. There are two factors that together cause this higher map resolution dependence of the PHANGS-ALMA data. 1.) Creating $N_{\rm{H}_2}$ maps using larger pixel sizes averages the column densities over larger regions. This leads to lower mean observed column densities, as very high column densities are usually detected at GMC scales, which are smaller than 1kpc \citep[][]{Leroy_2021}. This effect is especially apparent at the high column density end, as in the 1 kpc map resolution data column densities above log($N_{\rm{H}_2} / \rm{cm}^{-2})\sim 22.8$ are diluted by this effect. 2.) Observational data are limited by their sensitivity and completeness. Coarser resolution data have a reduced noise and are thus more complete \citep[see the comparison of native resolution vs. 150 pc map resolution data in PHANGS-ALMA,][]{Leroy_2021}. Therefore the coarser map resolution observations are sensitive to lower column densities compared to finer resolution observations. This effect is especially significant below log($N_{\rm{H}_2} / \rm{cm}^{-2})\sim 22.5$ in Fig. \ref{fig:f_N_150_1000pc} (right). TNG100 does not suffer from these sensitivity and incompleteness effects at lower column densities and therefore the map resolution dependence is less drastic. TNG100 is only affected by beam smearing. The core of the distribution [log($N_{\rm{H}_2} / \rm{cm}^{-2}) = 21 \; \rm{to} \; 22$] is robust to resolution effects in both simulations and observations.

\begin{figure*}
  \includegraphics[width=1.0\textwidth]{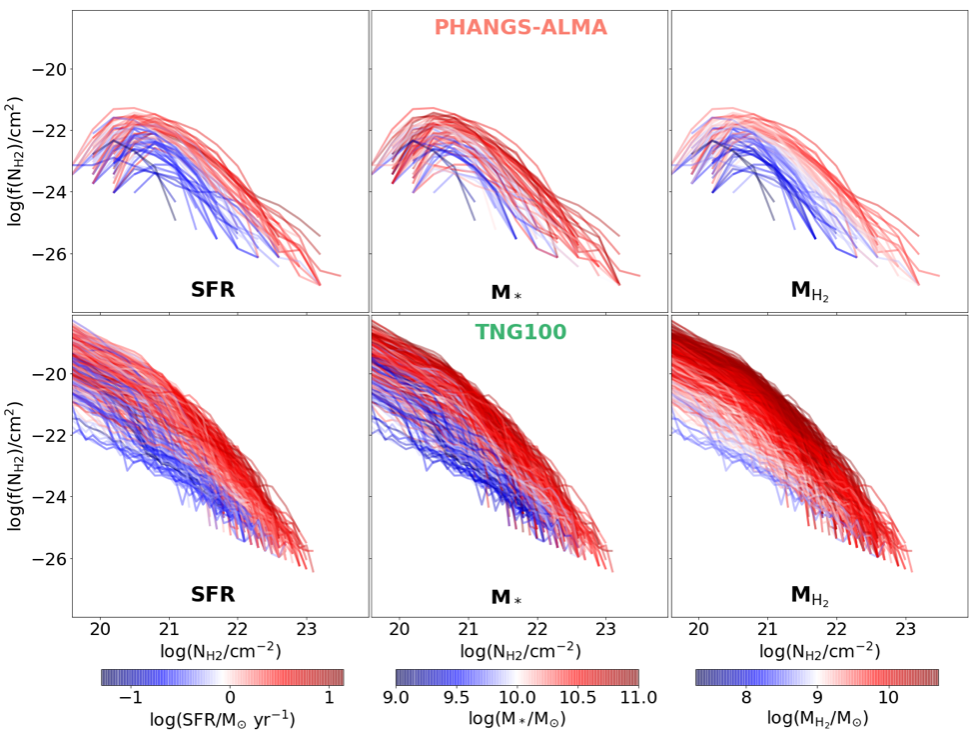}
  \caption{H$_2$ column density distributions ($f(N_{\rm{H_{2}}}$)) of individual PHANGS-ALMA and TNG100 ($z=0$) galaxies. The plots display the correlation of the individual $f(N_{\rm{H_{2}}}$) on the integrated SFR, $M_*$ and $M_{\rm{H}_2}$ of the galaxies. The flattening and steep drop of the PHANGS-ALMA $f(N_{\rm{H_{2}}}$) at lower column densities is due to the incompleteness and sensitivity limit of the observations. The individual $f(N_{\rm{H_{2}}}$) in both PHANGS-ALMA and TNG100 have similar shapes. This is likely due to the sample consisting of main-sequence star-forming galaxies. These galaxies mostly have rotating disks and are hypothesized to have} radially exponential gas profiles. The $f(N_{\rm{H_{2}}}$) correlate with integrated physical parameters (SFR, $M_*$ and $M_{\rm{H}_2}$) of the galaxies. The higher these parameters, the larger the galaxies, leading to a higher normalization of $f(N_{\rm{H_{2}}}$). Further, higher column densities are detected in more massive galaxies, implying that more dense gas is found in larger galaxies.
  \label{fig:f_N_PHANGS_individ}
\end{figure*}

\section{Does the H\texorpdfstring{$\bm{_2}$}{\_2} Column Density Distribution of Individual Galaxies depend on their Physical Properties?}

\label{sec:phangs_individ}

We study the $f(N_{\rm{H}_2}$) of individual objects in the PHANGS-ALMA survey and TNG100 ($z=0$) in order to explore how it depends on integrated physical properties of the galaxies. We calculate the individual $f(N_{\rm{H}_2}$) using Equation \ref{eq:f_N_H2}, but set the normalization parameters [$\Phi(x_i)$ and $w(x_i)$] equal to one. The individual column density distributions, colour-coded according to the integrated star formation rate (SFR), stellar mass ($M_*$) and H$_2$ mass ($M_{\rm{H}_2}$)\footnote{SFR and $M_*$ are taken from \href{https://sites.google.com/view/phangs/sample}{sites.google.com/view/phangs/sample}. $M_{\rm{H}_2}$ is calculated by summing up the surface density of individual pixels multiplied by pixel area (table found in datafileB1 at \href{https://www.canfar.net/storage/list/phangs/RELEASES/Sun_etal_2020b}{canfar.net/storage/list/phangs/RELEASES/Sun\_etal\_2020b})} of the corresponding galaxy are displayed in Fig. \ref{fig:f_N_PHANGS_individ}. For these calculations we use column density maps with a resolution of 150 pc. The colour coding of the plots reveals a connection between $f(N_{\rm{H}_2}$) and the physical parameters mentioned. We note that the integrated molecular masses of galaxies in TNG100 are generally higher compared to PHANGS-ALMA. This is to be expected, as TNG100 probes the full disk and is not limited by observational sensitivity and incompleteness limits when compared to the PHANGS-ALMA sample. Further, \cite{Leroy_2021} estimates that on average $\sim30$ per cent of molecular gas is missed by PHANGS-ALMA due to the limited field of view when compared to WISE3 luminosities. Additionally, TNG100 is possibly overestimating H$_2$ within the simulation at $z=0$ \citep[][]{Diemer_2019}.

The $f(N_{\rm{H}_2}$) of individual galaxies have very similar shapes in both observation and simulation. This is possibly related to the galaxies in the sample, which are main-sequence star-forming galaxies. These types of galaxies mostly have rotating disks and are hypothesized to have radially exponential gas profiles \citep{Leroy_2008, Stevens_2019}. While the diskiness and exponential gas profiles of the galaxies under consideration still need to be established, the similar $f(N_{\rm{H}_2}$) could indeed stem from similar gas profiles within these galaxies. In Sec. \ref{sec:red_evo} we further explore this possibility by comparing an analytical $f(N_{\rm{H}_2}$) model assuming radially exponential gas disks with simulated results.

While the shapes of the $f(N_{\rm{H}_2}$) are similar, the $f(N_{\rm{H}_2}$) also show a correlation with integrated physical parameters of the galaxies. The colour coding in Fig. \ref{fig:f_N_PHANGS_individ} indicates that the $f(N_{\rm{H}_2}$) are correlated with the integrated SFR, $M_*$, and $M_{\rm{H}_2}$ of the galaxies. The higher the SFR, $M_*$ and $M_{\rm{H}_2}$, the more massive these galaxies are, leading to a higher normalization of $f(N_{\rm{H}_2}$). Further, higher column densities are detected in more massive galaxies, implying that more dense gas is formed in larger galaxies. The higher abundance of denser gas in more massive galaxies could also lead to higher star formation rates, as more gas is found at densities suitable for star formation \citep[e.g. above log($N_{\rm{H}_2} / \rm{cm}^{-2}) \sim 21$,][]{Clark_2014}. This is possibly related to the correlation between the SFR surface density and molecular gas surface densities in galaxies \citep[e.g.][]{Bigiel_2008, Feldman_2020}. We however note that the correlation we find is related to the integrated SFR of the galaxy and not the SFR surface density. In the Appendix (Sec. \ref{app:cdd_dep}) we explore these correlations using the PHANGS-ALMA sample and provide a way to approximate $f(N_{\rm{H}_2}$) given physical parameters. It, however, remains unclear which galaxy property is the governing parameter for the shape of $f(N_{\rm{H}_2}$), as SFR, M$_*$ and M$_{\rm{H}_2}$ all have similar correlation strengths with parameters of the gamma distribution used to fit $f(N_{\rm{H}_2}$) in the Appendix.

\begin{figure*}
  \includegraphics[width=1.0\textwidth]{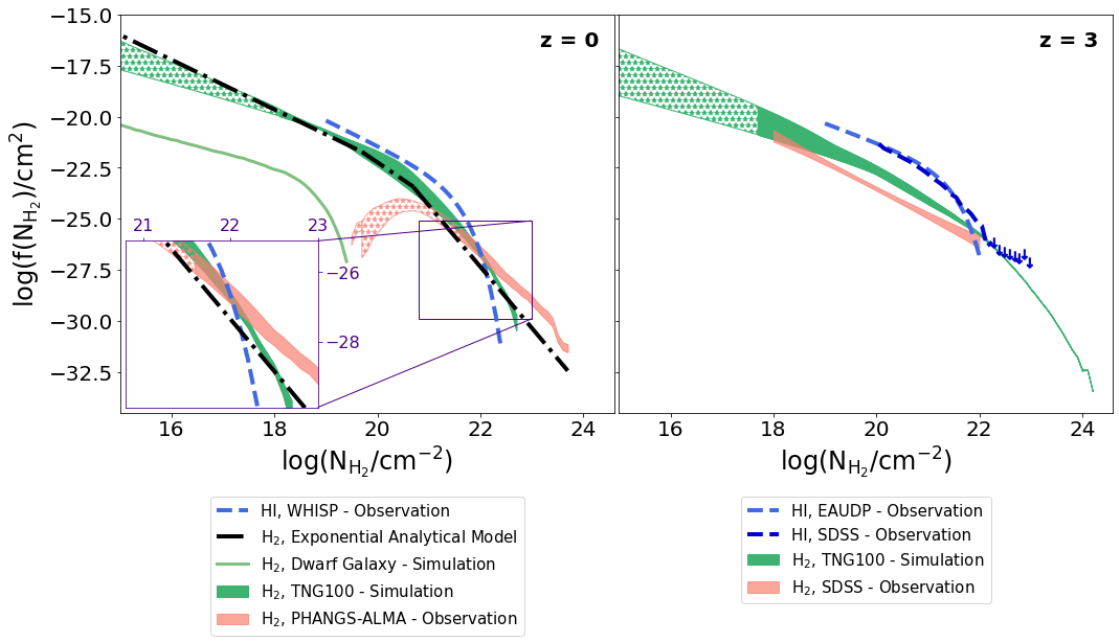}
  \caption{$f(N_{\rm{HI}})$ and $f(N_{\rm{H}_2})$ derived from both simulations and observations at $z=0$ and $z=3$. The column densities at which the $f(N_{\rm{H}_2})$ become unreliable due to incompleteness or simulation specifics are indicated by regions filled with the $\star$ symbol. 
  \textbf{Left:} $f(N_s)$ of TNG100, the PHANGS-ALMA survey (red band), the WHISP sample \citep[blue,][]{Zwaan_2005}, an analytical model \citep[black line,][]{Zwaan_2000} and the simulated dwarf galaxy from the GRIFFIN project (green line) at $z=0$. The $f(N_{\rm{H}_2})$ by TNG100 broadly reproduces the observations by the PHANGS-ALMA survey. The analytical $f(N_{\rm{H}_2})$ based on the assumption of an radially exponential gas profile in galaxies is a good approximation for $f(N_{\rm{H}_2})$ for both observations and simulations. The $f(N_{\rm{H}_2})$ derived from a simulated dwarf galaxy including a non-equilibrium chemical network displays similar slopes compared to TNG100. \textbf{Right:} $f(N_s)$ of TNG100 (green band), SDSS [\citep[red band,][]{Balashev_2018}, \citep[dark blue line,][]{Ho_2021}] and the EAUDP sample \citep[dark blue line,][]{Zafar_2013} at $z=3$. The $f(N_{\rm{H}_2})$ from TNG100 and the observational results based on the SDSS sample have matching slopes. The normalization between the two shows a $\sim 1$ dex difference, possibly arising due to differences in selection and environments probed.} 
  \label{fig:f_N_150pc_z0}
\end{figure*}

\section{The redshift evolution of the H\texorpdfstring{$\bm{_2}$}{\_2} column density distribution in simulations and observations}
\label{sec:red_evo}

We study $f(N_{\rm{H}_2})$ at $z=0$ and $z=3$ using both observations and simulations. First we study how recent observations compare to the state-of-the-art simulation TNG100 at both $z=0$ and $z=3$ and how the isolated dwarf galaxy simulation from the GRIFFIN Project fits into the column density distribution at $z=0$. Then we discuss the evolution of $f(N_{\rm{H}_2})$ from $z=0$ to $z=3$. Finally, we examine how $f(N_{\rm{H}_2})$ compares to $f(N_{\rm{HI}}$) at $z=0$ and $z=3$ to explore in which regions of galaxies the neutral atomic gas is dominating over the molecular gas.

\subsection{\texorpdfstring{$\bm{f(N_{\rm{H}_2})}$}{f(N\_H2)} at z=0}

In Fig. \ref{fig:f_N_150pc_z0} (left) the $f(N_{\rm{H}_2})$ from both observations, simulations and an analytical model at $z=0$ are displayed. For TNG100 we plot a band (green band) encompassing the three post-processing methods described in Sec. \ref{sec:TNG-100_data}. Note that below log($N_{\rm{H}_2}/\rm{cm}^{-2})$ $<$ 18 the post processing results for H$_2$ become unreliable as post-processing the simulations with different SPH kernel smoothing lengths leads to highly different results in that region. This region is represented by bands filled with the $\star$ symbol. The dwarf galaxy simulation $f(N_{\rm{H}_2})$ (green line) is based on the results from the on-the-fly chemical network included in the simulation. The red band encompasses the $f(N_{\rm{H}_2})$ from the PHANGS-ALMA survey using varying assumptions. It includes calculations using a stellar mass function based on the full sample and late-type galaxy only sample in \cite{Weigel_2016}. Further, we calculate the $f(N_{\rm{H}_2})$ with both a metallicity dependent $\alpha_{\rm{CO}}$ (see Sec. \ref{sec:PHANGS-ALMA_data}) and a constant $\alpha_{\rm{CO}} = 4.3 \; \rm{M}_{\odot} \; (\rm{K} \; \rm{km} \; \rm{s}^{-1} \; \rm{pc}^2)^{-1}$ \citep{Bolatto_2013}. We note that the drop of the PHANGS $f(N_{\rm{H}_2})$ at column densities below log($N_{\rm{H}_2} / \rm{cm}^{-2})\sim 21$ is not of physical origin, but due to the sensitivity and incompleteness limit of the observations, leading to the observations not probing the full disk. We also include an analytical model (black line) used to estimate $f(N)$ assuming radially exponential gas disks averaged over all possible inclinations \citep{Zwaan_2000} \footnote{With the $N_0$ parameter of the model, which determines the knee of the curve, set to $10^{20.7}$ cm$^{-2}$. For H$_2$ this is an ad-hoc choice.}. The analytical model is approximated by three linear functions.

\subsubsection{TNG100 broadly reproduces observations}

We study how TNG100 (green band, Fig. \ref{fig:f_N_150pc_z0} left) reproduces the $f(N_{\rm{H}_2})$ observed by the PHANGS-ALMA survey (red band, Fig. \ref{fig:f_N_150pc_z0} left). While there are differences in the $f(N_{\rm{H}_2})$, the observations are broadly reproduced by TNG100 in the column density ranges where simulation and observation specifics do not hinder a fair comparison.

One difference is that TNG100 does not reach as high a column density as observed by the PHANGS-ALMA survey. This is due to limitations of the simulation. Given the resolution of TNG100 gas densities that can be reached at given redshifts are limited (see Section \ref{sec:TNG50_100}). Further, for the regions below log($N_{\rm{H}_2}/\rm{cm}^{-2}) \sim 21.6$, TNG100 $f(N_{\rm{H}_2})$ shows a higher normalization than observed with PHANGS-ALMA. This can be explained by the sensitivity and incompleteness limit of the observations below those column densities. In the region between log($N_{\rm{H}_2}/\rm{cm}^{-2}) \sim 21.6 - 22.2$ both $f(N_{\rm{H}_2})$ are overlapping, albeit with TNG having a steeper slope when approximated as a linear function in log space ($\beta_{\rm{PHANGS}} \sim 2.3 - 2.35$, $\beta_{\rm{TNG}} \sim 3.15 - 3.6$). 

Given that the two $f(N_{\rm{H}_2})$ are based on vastly different methods of calculating the column densities (one being post-processed H$_2$ from a cosmological magnetohydrodynamical simulation, and one observations of CO(2--1), which are converted to H$_2$) the similarity between the two $f(N_{\rm{H}_2})$ is remarkable. Nonetheless further tests and studies are needed to explore the inconsistencies between observations and simulations, especially at the high column density end of $f(N_{\rm{H}_2})$. Higher resolution simulations would likely extend the $f(N_{\rm{H}_2})$ to higher column densities. Additionally, alternatives to the \cite{Springel_2003} sub-grid star formation prescription in order to resolve the cold gas phase in the simulations might be needed to reach the high column densities detected in observations. Including non-equilibrium chemistry in these simulations could also give more accurate representations of H$_2$ in the simulations \citep[e.g.][]{Maio_2022}. Finally, deeper observations would enable fair comparisons of $f(N_{\rm{H}_2})$ below log($N_{\rm{H}_2}/\rm{cm}^{-2}) \sim 21.6$.

\subsubsection{An analytical model closely matching TNG100}

We compare the analytical model by \cite{Zwaan_2000} (black line, Fig. \ref{fig:f_N_150pc_z0} left) with the results from TNG100 (green band, Fig. \ref{fig:f_N_150pc_z0} left) at $z=0$ to study how well the simulated predictions match the analytical model. We approximate the analytical model using three ad-hoc linear functions in the following H$_2$ column density ranges: log($N_{\rm{H}_2} / \rm{cm}^{-2}) \leq 20$, log($N_{\rm{H}_2} / \rm{cm}^{-2}) = 20 - 21$, log($N_{\rm{H}_2} / \rm{cm}^{-2}) \geq 21$ and therefore use these regions for a comparison. Generally, the analytical model assuming radially exponential gas disks in galaxies produces comparable results to calculating $f(N_{\rm{H}_2})$ using the post processed H$_2$ column densities of galaxies within TNG100. 

Below log($N_{\rm{H}_2} / \rm{cm}^{-2}) \leq 20$ the normalizations are comparable, but the TNG100 $f(N_{\rm{H}_2})$, depending on the post-processing prescription used, has a slightly lower slope compared to the analytical model ($\beta_{\rm{TNG}} \sim -0.75$ to $-1.15$, $\beta_{\rm{ana}} \sim -1.22$). In the range of log($N_{\rm{H}_2} / \rm{cm}^{-2}) = 20 - 21$ we find similar results, with the normalization matching, but a slightly higher slope in TNG100 ($\beta_{\rm{TNG}} \sim -1.94$ to $-2.05$, $\beta_{\rm{ana}} \sim -1.71$). Further, approximating the slope of TNG100 at log($N_{\rm{H}_2} / \rm{cm}^{-2}) \geq 21$ using a linear function in log space leads to similar results, with TNG100 producing higher slopes for $f(N_{\rm{H}_2})$ in this region ($\beta_{\rm{TNG}} \sim -3.4$ to $-3.8$, $\beta_{\rm{ana}} \sim -3.0$). Finally, the analytical model predicts slightly less systems in this column density range compared to TNG100. We note that \cite{Zwaan_2000} also proposes an analytical model based on Gaussian gas profiles. This model results in slopes of $\beta_{\rm{ana,gauss}} \sim -1$ at column densities below log($N_{\rm{H}_2} / \rm{cm}^{-2}) = 20.7$  and $\beta_{\rm{ana,gauss}} \sim -3$ at column densities above this threshold.

We conclude that while TNG100 produces $f(N_{\rm{H}_2})$ with slightly higher slopes and in some parts different normalizations, the $f(N_{\rm{H}_2})$ derived from the analytical model is still comparable and a good approximation. Since radially exponential gas disks are also a good approximation for disk galaxies in TNG100 \citep[e.g. HI disks described in][]{Stevens_2019} it appears natural that an analytical model making the assumption of radially exponential gas disks yields similar results. While the results of an analytical model using exponential gas disks matches predictions by TNG100 well, a Gaussian distribution within gas disks of galaxies yields similar results. Therefore, further studies of the distribution in gas disks and their relevance to $f(N_{\rm{H}_2})$ are required to fully understand how the gas disk distribution and $f(N_{\rm{H}_2})$ relate.

\subsubsection{A dwarf galaxy simulation producing similar slopes compared to TNG100}

We compare the $f(N_{\rm{H}_2})$ of the simulated dwarf galaxy from the GRIFFIN project, which includes a non-equilibrium chemical network tracking H$_2$ on the fly (green line, Fig. \ref{fig:f_N_150pc_z0} left) with the results of TNG100 (green band, Fig. \ref{fig:f_N_150pc_z0} left) at $z=0$. This helps us understand the impact for $f(N_{\rm{H}_2})$ when running simulations at sub-pc resolution including a non-equilibrium chemical network in an isolated environment. 

The $f(N_{\rm{H}_2})$ only probes one galaxy with a stellar mass of log($M_* / \rm{M}_{\odot}) \sim 7.3$\footnote{The stellar mass does not evolve much over the course of the simulation}. This leads to a number of differences when compared to a sample of galaxies. Due to the limited mass and size, the dwarf galaxy in the simulation only reaches column densities up to log($N_{\rm{H}_2} / \rm{cm}^{-2}) \sim 19.5$. 
The slope of both $f(N_{\rm{H}_2})$ is consistent. For the dwarf simulation the logarithmic slope is $\beta_{\rm{dwarf}} \sim -0.7$ before the drop off at log($N_{\rm{H}_2} / \rm{cm}^{-2}) \sim 18$. The slope found in TNG100 at those column densities is $\beta_{\rm{TNG}} \sim -0.7$ to $-1.1$. It is surprising that the slope of $f(N_{\rm{H}_2})$ of a single galaxy is so similar the slope of a large sample of galaxies with varying sizes, especially given the different methods for deriving molecular gas in these simulations. While it is difficult to disentangle the effects that the different galaxy properties and derivation methods of molecular gas have on the slope of $f(N_{\rm{H}_2})$, this is possibly a first indication that the slope $f(N_{\rm{H}_2})$ is not affected by non-equilibrium chemistry. Especially since the slope of $f(N_{\rm{H}_2})$ for individual main-sequence star-forming galaxies in TNG100 is similar below log($N_{\rm{H}_2} / \rm{cm}^{-2}) \lesssim 20$ and not majorly affected by galaxy properties. In order to further our understanding of how non-equilibrium chemistry might affect $f(N_{\rm{H}_2})$ a larger sample size of highly resolved simulated galaxies spanning a wider range of stellar masses would be needed. Alternatively, running and comparing the dwarf galaxy simulation by GRIFFIN without non-equilibrium chemistry with the current GRIFFIN model would also help disentangling the effects that non-equilibrum chemistry and galaxy properties have on the slope of $f(N_{\rm{H}_2})$. This is an interesting avenue to explore in the future.

\subsubsection{Which column densities contribute most to the H$_2$ mass density at $z=0$?}

As a final analysis of $f(N_{\rm{H}_2})$ at $z=0$ we study which column densities contribute the most to the overall mass density ($\rho_{\rm{mol}}$) in both TNG100 and the PHANGS-ALMA survey. Disentangling which column densities contribute the most to the mass density helps us understand in which regions of galaxies (e.g. the ISM, CGM, molecular clouds) most of the molecular gas is detected. Further, we can interpret if most of the gas is in regions suitable for star formation or not.

\begin{figure*}
  \includegraphics[width=1.0\textwidth]{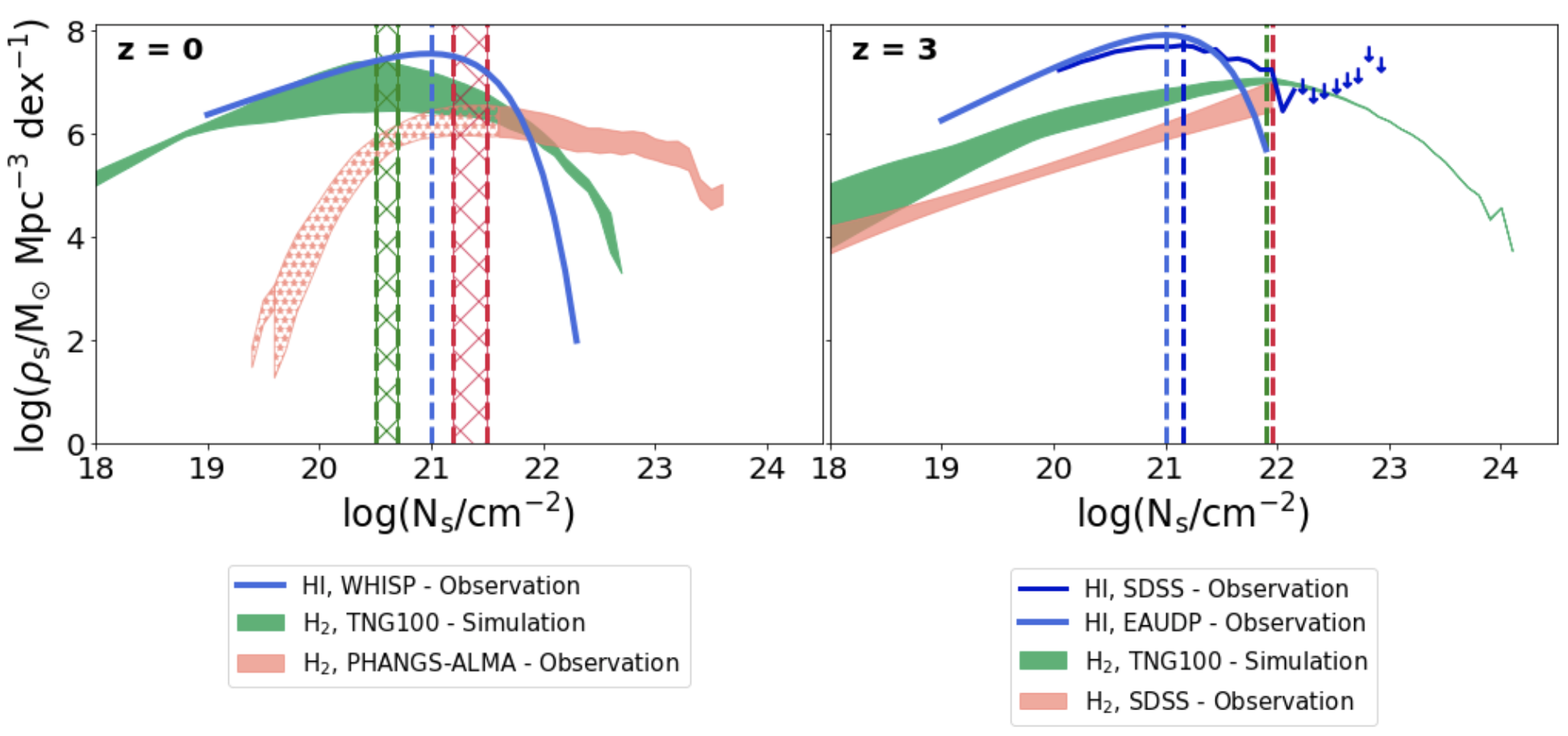}
  \caption{The mass density contribution per dex column density ($\rho_s$) of HI and H$_2$ derived from observations and simulations. Overall at $z=3$ the highest mass contribution of H$_2$ comes from denser gas compared to $z=0$ in both simulations and observations. Further, the mass density distributions suggest that HI dominates over H$_2$ at most column densities, making it an important contributor to the cold gas mass density of galaxies. H$_2$ starts dominating compared to HI at column densities above log($N_{\rm{H}_2} / \rm{cm}^{-2}) \sim 21.8-22$ at both redshifts. \textbf{Left:} Mass densities of H$_2$ and HI against column densities at $z = 0$ for TNG100 (green band), the PHANGS-ALMA survey (red band) and the WHISP sample \citep[blue line,][]{Zwaan_2005}. The H$_2$ highest mass density contribution can be constrained between log($N_{\rm{H}_2} / \rm{cm}^{-2}) \sim 20.5 - 21.5$ (column density regions typical for the ISM and, in part, molecular clouds)  \textbf{Right:} Mass densities of H$_2$ and HI against column densities at $z=3$ for TNG100 (green band), SDSS [\citep[red band,][]{Balashev_2018}, \citep[dark blue line,][]{Ho_2021}] and the EAUDP sample \citep[dark blue line,][]{Zafar_2013}. The highest H$_2$ mass density contribution can be constrained at log($N_{\rm{H}_2} / \rm{cm}^{-2}) \sim 22$ (column densities typical for molecular clouds)}
  \label{fig:mass_dens_z0}
\end{figure*}

In Fig. \ref{fig:mass_dens_z0} (left panel) we plot the mass densities as a function of H$_2$ column density. The red band corresponds to the PHANGS-ALMA results and the green band to the TNG100 results. For TNG100 the highest mass density contribution stems from column densities in the range log($N_{\rm{H}_2} / \rm{cm}^{-2}) \sim 20.5 - 20.7$. Therefore the majority of molecular gas in TNG100 is found at column densities typical for the ISM of galaxies, but below densities of molecular clouds \citep[e.g.][]{Spilker_2021} as opposed to less dense and diffuse regions surrounding galaxies, like the CGM.

Using numerical models \cite{Clark_2014} predict that star formation is possible in regions where the mean area averaged column density exceeds log($N_{\rm{H}_2} / \rm{cm}^{-2}) \sim 21$. TNG100 predicts the mass density peak slightly below the star formation threshold advocated by \cite{Clark_2014} and therefore in a region not suitable for star formation. This fraction of the gas could be either in regions where the molecular gas has been depleted due to star formation, or in regions that are possibly in the process of collapsing into denser regions.

In PHANGS-ALMA we find an overall flatter distribution of the H$_2$ mass densities in the regions where the observations are complete and when compared to TNG100. The highest contribution to the overall mass density is in the range of log($N_{\rm{H}_2} / \rm{cm}^{-2}) \sim 21.2 - 21.5$. This is at densities detected in the ISM and typical for molecular clouds. We note that for the 1 kpc resolution PHANGS-ALMA data the highest contribution shifts to log($N_{\rm{H}_2} / \rm{cm}^{-2}) \sim 21$. However, it is not trivial to quantify how much this is an effect of higher completeness at lower resolutions compared to averaging over a larger area.

The mass density peak in PHANGS-ALMA is detected at densities above the star formation threshold advocated by \cite{Clark_2014}. While this is inconsistent with results by TNG100, we note that the observations of PHANGS-ALMA are incomplete in this region. It is therefore conceivable that deeper observations of molecular gas in these galaxies may shift the observed column density contributions to lower column densities.

In conclusion, when combining results by observations and simulations, the highest H$_2$ mass density contribution is found at column densities detected within the ISM of galaxies and partly in regions observed in local molecular clouds.

\subsection{\texorpdfstring{$\bm{f(N_{\rm{H}_2})}$}{f(N\_H2)} at z=3}

\subsubsection{TNG100 broadly reproducing observations}

Here we compare the $f(N_{\rm{H}_2})$ based on composite SDSS H$_2$ absorption spectra \citep[][]{Balashev_2018} to the results of TNG100 at $z=3$. In Fig. \ref{fig:f_N_150pc_z0} (right plot) we display these $f(N_{\rm{H}_2})$. The data from \cite{Balashev_2018} (red band) only includes column densities of log($N_{\rm{H}_2}/\rm{cm}^{-2})$ = 18 - 22 and we therefore can only compare the slopes before the steeper drop of $f(N_{\rm{H}_2})$ at higher column densities. Both $f(N_{\rm{H}_2})$ have similar slopes in this region ($\beta_{\rm{balashev}} \sim 1.13 - 1.45$ ,  $\beta_{\rm{TNG}} \sim 0.88 - 1.47$ ). The normalization of the observed $f(N_{\rm{H}_2})$ is $\sim$ 1 dex lower than that predicted by TNG100 in most column density regions.

This may be caused by the different methods of deriving $f(N_{\rm{H}_2})$. The observed $f(N_{\rm{H}_2})$ is based on absorption line studies of DLAs, which are typically observed at high impact parameters surrounding galaxies \citep[e.g.][]{Peroux_2011, Christensen_2014, Krogager_2017}, while TNG100 relies on post-processed H$_2$ column density maps, which include all regions of galaxies. Further, these observations might be biased towards galaxies in group environments \citep[][]{Hamanowicz_2020}. Given this, while there are still inconsistencies between simulation and observation, the two $f(N_{\rm{H}_2})$ are remarkably close in slope. Further studies, including high spatial resolution molecular gas observations or post-processing TNG100 using ray-casting codes at typical impact parameters of absorption line systems, might help alleviate some of these inconsistencies and are an interesting avenue for future studies.

We stress that another model by \cite{Krogager_2020} using the fraction of cold gas absorption in strong HI selected absorbers derived by \cite{Balashev_2018} predicts an $f(N_{\rm{H}_2})$ with a knee at log($N_{\rm{H}_2}/\rm{cm}^{-2})$) $\sim 21$ and a highest column density of log($N_{\rm{H}_2}/\rm{cm}^{-2})$ $\sim 23$. While the model is also using the \cite{Blitz_2006} method for splitting the cold gas into a neutral and molecular fraction as for TNG100, the results are inconsistent with the predictions made by TNG100, which estimates the knee of  $f(N_{\rm{H}_2})$ to occur at log($N_{\rm{H}_2}/\rm{cm}^{-2})$ $\sim 22$ and includes column densities beyond log($N_{\rm{H}_2}/\rm{cm}^{-2})$ $= 24$.

In conclusion, both observations and simulations have $f(N_{\rm{H}_2})$ with well matching slopes in the overlapping regions. However, they differ in normalization by $\sim 1$ dex. Thus, the results by observations and simulations are in tension for the overlapping H$_2$ column density regions at $z=3$.

\subsubsection{Which column densities contribute most to the H$_2$ mass density at $z=3$?}

As a final analysis of $f(N_{\rm{H}_2})$ at $z=3$ we study which column densities contribute the most to the overall mass density ($\rho_{\rm{mol}}$) in both TNG100 and the $f(N_{\rm{H}_2})$ by \cite{Balashev_2018} derived from composite H$_2$ spectra.

In Fig \ref{fig:mass_dens_z0} (right panel) we plot the mass densities for each H$_2$ column density. The peak of the H$_2$ mass density is not reached by the \cite{Balashev_2018} data (red band), meaning that we can only set a limit of log($N_{\rm{H}_2} / \rm{cm}^{-2}) \gtrsim 22$. This is in regions typically observed within molecular clouds. Further, it is well above the density threshold for star formation. The TNG100 results show that the highest mass density contribution is at densities of log($N_{\rm{H}_2} / \rm{cm}^{-2}) \sim 21.9$, so slightly below the limit that one can set with observations.

\subsection{Denser molecular gas found at high redshifts}

In this section we study how $f(N_{\rm{H}_2})$ evolves from redshift $z=0$ to $z=3$. The $f(N_{\rm{H}_2})$ for both redshifts is shown in Fig. \ref{fig:f_N_150pc_z0}.

In TNG100 (green bands) the slopes below log($N_{\rm{H}_2} / \rm{cm}^{-2}) = 20$ are  similar ($\beta_{\rm{TNG,z=0}} \sim 0.75-1.14$, $\beta_{\rm{TNG,z=3}} \sim 0.73-1.10$) and show little to no evolution. At column densities above that differences start to arise. At $z=0$, TNG100 predicts molecular gas up to column densities of log($N_{\rm{H}_2} / \rm{cm}^{-2}) \sim 23$. At $z$=3 column densities beyond log($N_{\rm{H}_2} / \rm{cm}^{-2}) \gtrsim 24$ are reached in TNG100. This indicates that denser H$_2$ gas exists in the earlier Universe. It is the case that physical densities are intrinsically higher in the high-redshift versus low-redshift Universe. At the same time, this prediction from TNG100 could be affected by its finite numerical resolution. Further, there is a steeper drop off at high column densities at $z=0$ compared to $z=3$ in TNG100, where the $f(N_{\rm{H}_2})$ is flatter at high column densities. 
Due to limitations in the observations, we cannot make similar statements at the high column densities using observations. We, however, find that in the overlapping region $f(N_{\rm{H}_2})$ of both the SDSS sample \citep{Balashev_2018} and the PHANGS-ALMA survey are similar. The $f(N_{\rm{H}_2})$  of \cite{Balashev_2018} is a good continuation of the $f(N_{\rm{H}_2})$ found in the PHANGS-ALMA survey. We therefore expect larger differences in the $f(N_{\rm{H}_2})$ to arise at higher column densities. This would mean that the largest differences of $f(N_{\rm{H}_2})$ arise at the densest molecular regions in the Universe. Observations at $z=3$ with higher column densities are needed in order to test if the predictions by TNG100 are correct.

Figure \ref{fig:mass_dens_z0} shows that the column densities contributing the most to the molecular gas mass densities are shifting towards higher column densities at $z=3$. When combining the results from observations and simulations we find the following: While at $z=0$ the highest contribution is found at column densities of log($N_{\rm{H}_2} / \rm{cm}^{-2}) \sim 20.5 - 21.5$, at $z=3$ it is found at column densities of log($N_{\rm{H}_2} / \rm{cm}^{-2}) \sim 21.9 - 22$.
When assuming that the column density relates to the density of the gas, denser gas found at higher redshifts is in line with observations of the star formation rate across cosmic time, which is higher at $z=3$ compared to $z=0$ \citep[][]{Madau_2014, Tacconi_2020}. The shape of the cosmic molecular mass density as a function of redshift is similar to the shape of the SFR density, making a coupling of these two quantities likely. Therefore, one would expect that more molecular gas found in denser regions leads to a higher global star formation rate in galaxies \citep[][]{Peroux_2020}. When assuming that the column density relates to the density of the gas, this is exactly what we observe when studying the column density distributions at $z=0$ and $z=3$.

\subsection{Is H\texorpdfstring{$\bm{_2}$}{\_2} dominating the higher column densities?}

We compare $f(N_{\rm{H}_2})$ and $f(N_{\rm{HI}})$ at $z=0$ and $z=3$ to study the column densities at which H$_2$ overtakes HI. In the following sections we compare these derived densities with the combined results of $f(N_{\rm{H}_2})$ derived from both observation and simulation. At $z=0$ we compare the $f(N_{\rm{H}_2})$ with $f(N_{\rm{HI}})$ derived by \cite{Zwaan_2005}. At $z=3$ we compare the $f(N_{\rm{H}_2})$ with the $f(N_{\rm{HI}})$ derived by \cite{Zafar_2013} and \cite{Ho_2021}.

\subsubsection{HI and H$_2$ Column Density Distributions at $z = 0$}

\label{sec:mass_dens_z0}

The $f(N_{\rm{HI}})$ at $z=0$ from \cite{Zwaan_2005} is based on 21-cm maps of 355 galaxies of the WHISP sample \citep[][]{van_der_Hulst_2001}. The WHISP sample covers galaxies of all Hubble types from S0 to Im and a considerable luminosity range and were selected using the Uppsala General Catalogue (UGC) of galaxies \citep[][]{Nilson_1973}. The median spatial resolution reached by these observation is $\sim 1.4$ kpc.

The $f(N_s)$ (left panel in Fig. \ref{fig:f_N_150pc_z0}) and $\rho_s$ (left panel in Fig \ref{fig:mass_dens_z0}) at $z = 0$ show that H$_2$ starts to dominate the mass density at column densities above log($N_{\rm{H}_2} / \rm{cm}^{-2}) \sim 21.8 - 22$ \footnote{We note that the TNG100 $\rho_{\rm{H}_2}$ band implies that at $z=0$ the H$_2$ mass density is roughly equal in the log(${N_{\rm{H}_2} / \rm{cm}^{-2}) \sim 19 - 20.5}$ column density region. We attribute this to a possible over-prediction of H$_2$ (and HI) in the simulation compared to observations at $z=0$ \citep[][]{Diemer_2019}. Deeper observations are needed to quantify how high the contribution of molecular gas is at these densities.}. This is consistent with results from \cite{Schaye_2001}, who predicted that HI clouds with $N_{\rm{HI}} \gtrsim 10^{22} \rm{cm}^{-2}$ transform to molecular clouds before reaching higher column densities. Similar predictions have also been made more recently by \cite{Altay_2011} and \cite{Bird_2014} using (magneto-)hydrodynamical simulations. 

These results imply that while molecular gas dominates the high column densities above log($N_s / \rm{cm}^{-2}) \gtrsim 22$, HI dominates the majority of the column density regions found within the interstellar medium (including column density regimes typical for molecular clouds), making neutral gas an important contributor to the cold gas mass found within galaxies at $z = 0$.

\subsubsection{HI and H$_2$ Column Density Distributions at $z = 3$}

The two $f(N_{\rm{HI}})$ at $z \sim 3$ are based on HI-absorption systems (sub-DLAs and DLAs). The calculation therefore relies on pencil beam observations of HI-column densities as studying 21-cm HI in emission is not feasible at this redshift. The $f(N_{\rm{HI}})$ by \cite{Ho_2021} is based on the Sloan Digital Sky Survey Data Release 16 which were analyzed using Gaussian processes, where DLAs are detected using Bayesian model selection. While SDSS-DR16 includes redshifts between $z = 2$ and $z = 5$, we only use the results of the $z = 2.5 - 3$ integration for our comparison. The $f(N_{\rm{HI}})$ by \cite{Zafar_2013} is based on the ESO UVES advanced data products (EUADP) sample and includes measurements in the $z \sim 1.5 - 3.1$ range. The $f(N_{\rm{HI}})$ of both samples show comparable results up to log($N_{\rm{HI}}/\rm{cm}^{-2}) \sim$ 22. Above this density SDSS results display a possible flattening of the $f(N_{\rm{HI}})$. This flattening would be inconsistent with the predictions of the maximum $N_{\rm{HI}}$ by \cite{Schaye_2001}, but the Gaussian process analysis shows that the $f(N_{\rm{HI}})$ in that region is also consistent with 0 and therefore not well constrained. We further note that while the SDSS-DR16 sample is larger than the EAUDP sample, the resolution is lower. The lower resolution could lead to blending at higher column densities, which would lead to measurements of column densities above log($N_{\rm{HI}}/\rm{cm}^{-2}) \sim$ 22.

The $f(N_s)$ (right panel in Fig. \ref{fig:f_N_150pc_z0}) and $\rho_s$ (right plot in Fig \ref{fig:mass_dens_z0}) at $z = 3$ show that H$_2$ starts to dominate the mass density at column densities between log($N_{\rm{H}_2} / \rm{cm}^{-2}) \sim 21.5-22$. As for $z = 0$, neutral gas is an important contributor to the global mass in a wide range of regions found in the ISM including higher density regions typical of molecular clouds.

\subsubsection{HI - An Important Contributor to the Cold Gas Mass of Galaxies}

In conclusion, Figure \ref{fig:mass_dens_z0} indicates that HI dominates over H$_2$ at most column densities. The HI column density contributing most to the overall mass density (blue vertical lines) has a higher mass contribution than H$_2$ at both redshifts. HI could therefore be an important contributor to the cold gas mass of galaxies at $z=0$ and $z=3$. 

The column density contributing the most to the overall HI gas mass density is at log(${N_{\rm{HI}} / \rm{cm}^{-2}) \sim 21}$ for both redshifts. In contrary the highest contributing column density of H$_2$ evolves with redshift. It is log(${N_{\rm{H}_2} / \rm{cm}^{-2}) \sim 22}$ at $z=3$ and less than log(${N_{\rm{H}_2} / \rm{cm}^{-2}) \sim 21.5}$ at $z=0$. We note that the molecular phase of the gas cycle is likely to be shorter than the neutral atomic phase as indicated by cold gas depletion time scales \citep[][]{Peroux_2020}. Therefore the molecular gas phase is more dynamic and variations in the gas densities are to be expected across cosmic time.

The HI column density contributing the most to the HI mass density is log(${N_{\rm{HI}} / \rm{cm}^{-2}) \sim 21}$. These high column densities are not found in diffuse gas (e.g. the CGM), but are typical of column densities found the ISM. 

At both $z=0$ and $z=3$ H$_2$ starts to dominate the mass density at column densities in the log($N_{\rm{H}_2} / \rm{cm}^{-2}) \sim 22$ range therefore showing little to no evolution of this observable. This is consistent with the predictions made by \cite{Schaye_2001} suggesting that little to no gas is found in the neutral phase at column densities above log($N_{\rm{HI}}/\rm{cm}^{-2}) \gtrsim$ 22 due to the clouds turning molecular at those column densities. 

\section{Discussion}

\label{sec:discussion}

Given the evolution of the H$_2$ comoving mass density over cosmic time \citep[e.g.][]{Riechers_2019, Peroux_2020, Decarli_2020} , changes in the normalisation or shape of $f(N_{\rm{H}_2})$ are expected. The $f(N_{\rm{H}_2})$ derived from both observations and simulations corroborate this hypothesis with various changes of the $f(N_{\rm{H}_2})$ across cosmic time. In general, the combined results of observations and simulations imply that molecular gas is more often found in systems of higher column densities at $z=3$ when compared to $z=0$. These changes in the $f(N_{\rm{H}_2})$ are in line with the higher comoving molecular mass densities detected at $z=3$. Combined with the higher star formation rate density detected around cosmic noon \citep[e.g.][]{Madau_2014} the results imply that the overall denser molecular gas at higher redshifts lead to a higher global star formation rate. While we study global properties in this work, these results are similar to findings of local observations of nearby star-forming galaxies where a correlation between the SFR surface density and H$_2$ surface density is well established \citep[e.g. the molecular Schmidt law in][]{Bigiel_2008}.

\cite{Rahmati_2013} have demonstrated that observed $f(N_{\rm{HI}})$ can be accurately reproduced using the cosmological hydrodynamical simulation EAGLE \citep[][]{Schaye_2015}. Similarly, at $z=3$, the cosmological simulation Illustris \citep[][]{Genel_2014, Vogelsberger_2014, Vogelsberger_2014b, Sijacki_2015} reproduces $f(N_{\rm{HI}})$ of observations \citep[][]{Noterdaeme_2009, Zafar_2013, Prochaska_2010} accurately. However, there are still tensions between Illustris and observations below $z=3$ \citep[][]{Bird_2014}. However, \cite{Villaescusa-Navarro_2018} demonstrate that these tensions are not apparent in the successor of Illustris. Comparing results by TNG100 of the IllustrisTNG project with observations \cite{Villaescusa-Navarro_2018} find that $f(N_{\rm{HI}})$ is accurately reproduced at $z \lesssim 5$.

While it has been demonstrated that $f(N_{\rm{HI}})$ is consistent with observations in different (magneto-)hydrodynamical cosmological simulations, there are still a number of inconsistencies for $f(N_{\rm{H}_2})$, despite the broad similarities of simulated and observed $f(N_{\rm{H}_2})$. At $z=0$, \cite{Klitsch_2019} demonstrate that TNG100 predict more low column density molecular gas compared to constraints by the ALMACAL survey \citep[e.g.][]{Otteo_2016, Bonato_2018, Klitsch_2018} and, similarly to this work, does not reach the high column densities detected in observations \citep[][]{Zwaan_2006}. These short-comings are, in part, due to TNG100 not resolving the cold gas phase of the ISM. These simulation specifics stem from limitations in resolution and sub-grid star formation models. Further, at $z=0$ TNG100 might over predict H$_2$ compared to observational findings \citep[][]{Diemer_2019}, especially when not taking observational apertures into account \citep[][]{Popping_2019}.

At $z=3$, we find a $\sim 1$ dex, difference in normalization for $f(N_{\rm{H}_2})$, which could arise due to the difference in selection and environments probed. The observational $f(N_{\rm{H}_2})$ at $z=3$ is based on DLA studies. DLAs mostly trace the outskirts of galaxies \citep[e.g.][]{Peroux_2011, Christensen_2014, Krogager_2017} and are often associated with group environments \citep[][]{Hamanowicz_2020}, while in TNG100 the full disk with no constraints on the environment of the galaxies is probed. Therefore, further efforts, on both the observational and simulation side are needed. On the simulation side more accurate representations of the cold gas phase is needed, including different sub-grid models of star formation, higher resolution and the inclusion of non-equilibrium chemistry. On the observational side we need better constraints of $f(N_{\rm{H}_2})$, especially at $z=3$. Preferably, this could be achieved by a combination of high spatial resolution galaxy observations and a larger sample of H$_2$ absorption line systems at $z=3$.

Non-equilibrium chemistry networks \citep[e.g.][]{Glover_2007b, Glover_2012, Gong_2017}, have recently been used to model the cold gas phase in simulations on the fly. Such models have been implemented in simulations of individual regions of galactic disks \citep[e.g.][]{Walch_2015, Rathjen_2021, Hu_2021}, isolated galaxies \citep[e.g.][]{Richings_2016, Hu_2016, Lahen_2019}, and more recently in cosmological simulations \citep{Maio_2022}. These studies have shown that non-equilibrium chemistry e.g. heavily influences the H$_2$ mass fraction at low metallicities \citep[][]{Hu_2021}, affect the chemical make-up of outflows \citep[][]{Richings_2016} and more accurately reproduce cosmological H$_2$ mass densities of observations \citep[][]{Maio_2022}. As a first attempt to study how and if non-equilibrium chemistry affects $f(N_{\rm{H}_2})$, we compare the time- and inclination-averaged $f(N_{\rm{H}_2})$ derived from a dwarf galaxy simulation by the GRIFFIN Project with the $f(N_{\rm{H}_2})$ derived by TNG100. The normalization of the dwarf galaxy $f(N_{\rm{H}_2})$ in the overlapping column density region is lower than for TNG100 $f(N_{\rm{H}_2})$ due to the highly different stellar masses that are probed. Interestingly, the slope of the $f(N_{\rm{H}_2})$ is similar, even though the samples and cold gas models are vastly different. We cannot disentangle the effects that non-equilibrium chemistry and the different samples have on $f(N_{\rm{H}_2})$ with our current study. Nonetheless, this could be a first indication that non-equilibrium chemistry might not affect the slope of $f(N_{\rm{H}_2})$, especially since the slope of $f(N_{\rm{H}_2})$ for individual main-sequence star-forming galaxies in TNG100 is similar below log($N_{\rm{H}_2} / \rm{cm}^{-2}) \lesssim 20$ and not majorly affected by galaxy properties. However, comparisons between simulation runs of the same galaxy with and without non-equilibrium chemistry could help understand if and how $f(N_{\rm{H}_2})$ is affected by non-equilibrium chemistry. Further, studies with larger samples, similar to \cite{Maio_2022}, are needed to further investigate how non-equilibrium chemistry might affect $f(N_{\rm{H}_2})$.

The global $f(N_{\rm{H}_2})$ and that of individual main-sequence star-forming galaxies give first indications that its shape could be related to the gas distribution within gas disks. Exponential gas distributions have not only been observed in disk galaxies \citep[e.g.][]{Leroy_2008}, but also reproduced in simulated ones \citep[e.g. in TNG100,][]{Stevens_2019}. An analytical model, based on exponential gas distribution in disks \citep[][]{Zwaan_2000} broadly reproduces $f(N_{\rm{H}_2})$ of simulations and observations and is giving a first indication that these two distributions are related. Nonetheless, analytical models with e.g. Gaussian gas distributions in gas disks yield similar results. Therefore, it currently remains unclear how closely coupled the shape of $f(N_{\rm{H}_2})$ and the gas distribution in gas disks are. Further studies are needed for a complete understanding to confirm the hypothesis of this connection between these two observables.

At $z=0$, observations have shown that neutral atomic hydrogen dominates the total mass of the neutral ISM, with $M_{\rm{HI}} \sim 2-10 \; M_{\rm{mol}}$ \citep[e.g.][]{Saintonge_2011, Saintonge_2022}. In studies at higher redshifts, it is often assumed that the neutral atomic component can be omitted and H$_2$ is assumed to be the dominant gas component in galaxies \citep[e.g. between z=0.4 and 4,][]{Tacconi_2018}. In part, this is due to technical limitations, as the HI 21cm emission line is not observable at higher redshifts with current instruments. Further, the molecular mass density peaks within this redshift range, while the neutral atomic mass density remains fairly constant across cosmic time, possibly making molecular gas an important contributor to the overall gas mass of galaxies within this redshift range (especially around cosmic noon). However, it still remains unclear what the contribution of the neutral atomic gas phase is to galaxies at higher redshifts. \cite{Heintz_2021} have given first indications of the contribution of HI at higher redshifts, by exploiting [CII] as a tracer for neutral atomic gas. The results indicate that at $z=4-6$ the contribution of HI is substantial, with the HI mass being equal to the dynamical mass of galaxies. At $z\sim2$ the contribution of HI is found to be less substantial, with the HI mass being between 0.2 - 1 dex lower than the dynamical mass of galaxies. Therefore, at $z\sim 2$, the contribution by molecular gas or the stellar component is possibly higher. Comparing $f(N_{\rm{H}_2})$ and $f(N_{\rm{HI}})$ we, however, find that HI is an important contributor to the overall cold gas mass found in the ISM of galaxies (see Section \ref{fig:mass_dens_z0}) at both redshift $z=0$ and $z=3$. We therefore caution from omitting the neutral atomic gas component in studies at these redshifts.

\section{Conclusions}

\label{sec:conclusions}

In this work we study the H$_2$ column density distribution [$f(N_{\rm{H}_2})$] at redshift $z=0$ and $z=3$ using observations and simulations. On the observational side we use data from the PHANGS-ALMA survey \citep[][]{Leroy_2021} at $z=0$ and from an H$_2$ absorption line study by \cite{Balashev_2018} at $z=3$ based on SDSS data. On the simulation side we use data from TNG100 of the IllustrisTNG project \citep{Marinacci_2018, Springel_2018, Naiman_2018,Nelson_2018, Pillepich_2018} at both redshift $z=0$ and $z=3$ and a high-resolution isolated dwarf galaxy simulation including a non-equilibrium chemical network by the GRIFFIN project \citep[][]{Lahen_2019, Lahen_2020, Lahen_2020b} meant to represent a low-redshift dwarf galaxy.

In summary our analysis includes the following studies: 

\begin{itemize}
    \item We study how the integrated properties of galaxies in the PHANGS-ALMA sample shape the $f(N_{\rm{H}_2})$ of individual objects.
    \item We contrast the $f(N_{\rm{H}_2})$ from observations and simulations to test how predictions made by TNG100 match observations.
    \item We study how well analytical models match results by TNG100.
    \item We compare results from a simulation including non-equilibrium chemistry (GRIFFIN Project) with results from the post-processed simulation TNG100.
    \item We study the evolution of $f(N_{\rm{H}_2})$ from $z=3$ to $z=0$.
    \item We explore which column densities contribute most to the overall H$_2$ and HI mass density at $z=0$ and $z=3$.
    \item We investigate how the $f(N_{\rm{H}_2})$ compare to $f(N_{\rm{HI}})$ based on the WHISP sample \citep[][]{Zwaan_2005}, EAUDP sample \citep{Zafar_2013} and SDSS data \citep{Ho_2021} to examine in which regions of galaxies molecular gas dominates over neutral atomic gas.
\end{itemize}

In conclusion our findings are the following:

\begin{itemize}
\setlength\itemsep{1em}
    \item The shapes of the ${f(N_{\rm{H}_2})}$ of individual galaxies in the PHANGS-ALMA and the TNG100 sample at ${z=0}$ are similar. This is possibly related to the galaxies in the sample. The sample consists of main-sequence star-forming galaxies, which typically have rotating disks and are hypothesized to have radially exponential gas profiles. \citep[][]{Leroy_2008, Stevens_2019}. The radially exponential gas profiles could potentially be the cause of the similar ${f(N_{\rm{H}_2})}$ observed for individual galaxies. Further, the normalization of ${f(N_{\rm{H}_2})}$ and highest observed H$_2$ column densities depend on the integrated star formation rate (SFR), stellar mass (${M_*}$) and H${_2}$ mass (${M_{\rm{H}_2}}$) of the galaxy. More massive galaxies lead to a higher normalization of the ${f(N_{\rm{H}_2})}$ of individual galaxies. The ${f(N_{\rm{H}_2})}$ indicates that more massive galaxies produce more dense gas.
    
    \item TNG100 broadly reproduces the ${f(N_{\rm{H}_2})}$ we observe at both ${z=0}$ and ${z=3}$, albeit with some key differences. At ${z=0}$ TNG100 produces steeper slopes for the ${f(N_{\rm{H}_2})}$ compared to PHANGS-ALMA. Further, observations detect column densities up to log(${N_{\rm{H}_2} / {cm}^{-2}) \sim 24}$ at ${z=0}$. $\:$ Such high column densities are not present in TNG100 at that redshift. This is potentially due to resolution effects and the star formation sub-grid interstellar medium model, both of which could inhibit the formation of high column densities of cold gas phases. At ${z=3}$ the normalization of the ${f(N_{\rm{H}_2})}$ is higher in the simulations compared to observations for the majority of the regions. This is likely due to the different environments probed by SDSS observations. Nonetheless, the slopes ${f(N_{\rm{H}_2})}$ in TNG100 and from observations are in good agreement at ${z=3}$.
    
    \item The dwarf galaxy simulation from the GRIFFIN project produces similar slopes as TNG100 for ${f(N_{\rm{H}_2})}$ in the overlapping column density region. It is surprising that the slope of ${f(N_{\rm{H}_2})}$ of a single simulated galaxy including a non-equilibrium chemistry network is so similar to the slope of a large sample of galaxies where H${_2}$ was derived using post-processing prescriptions. This could be a first indication that non-equilibrium chemistry might not majorly affect the slope of ${f(N_{\rm{H}_2})}$. However, further studies are needed to understand how and if non-equilibrium chemistry affects ${f(N_{\rm{H}_2})}$.
    
    \item {The slopes of ${f(N_{\rm{H}_2})}$ below log(${N_{\rm{H}_2} / \rm{cm}^{-2}) \sim 20}$ show little to no evolution from ${z=3}$ to ${z=0}$. As indicated by the  ${f(N_{\rm{H}_2})}$ derived from TNG100, we expect an evolution of the ${f(N_{\rm{H}_2})}$ to arise at higher column densities.}
    
    \item {The mass density distributions of the neutral atomic and molecular gas phase indicate that HI dominates over H$_{{2}}$ at most column densities and shows that HI could be an important contributor to the cold gas mass of galaxies at ${z=0}$ and ${z=3}$.}
    
    \item {The H$_2$ column density contributing most to the overall molecular gas density evolves with redshift. When combining data from observations and simulations, we find that the shift is from log(${N_{\rm{H}_2} / \rm{cm}^{-2}) \sim 20.5 - 21.5}$ at ${z=0}$ to log(${N_{\rm{H}_2} / \rm{cm}^{-2}) \sim 21.9 - 22}$ at ${z=3}$. We therefore find that more gas in denser regions is found at ${z=3}$ compared to ${z=0}$. These results are in line with observations of the star formation rate across cosmic time, which is higher at ${{z=3}}$ compared to ${z=0}$. The shape of the cosmic molecular mass density as a function of redshift is similar to the shape of the SFR density, making a coupling of these two quantities likely. Therefore, one would expect that more molecular gas found in denser regions leads to a higher global star formation rate of in galaxies \citep[][]{Peroux_2020}. When assuming that the column density relates to the density of the gas, this} is exactly what we observe when studying the column density distributions at both redshifts.
    
    \item {Contrary to H$_{{2}}$, the column density contributing most to the HI gas mass density [log(${N_{\rm{HI}} / \rm{cm}^{-2}) \sim 21}$] does not evolve with redshift. Given that the molecular phase of the gas cycle is likely to be shorter than the neutral atomic gas phase, more variations in the molecular gas densities are expected. The highest column density contribution of HI is therefore found in regions of the ISM, and not in more diffuse regions like the CGM.}
    
    \item {H$_2$ starts dominating compared to HI at column densities above log(${N_{\rm{H}_2} / \rm{cm}^{-2}) \sim 21.8 - 22}$ at both redshifts. This is consistent with results by \cite{Schaye_2001}, who predicted that HI clouds with log(${N_{\rm{HI}} / \rm{cm}^{-2}) \gtrsim 22}$ do not occur due to the clouds turning molecular before reaching higher column densities. Further, this implies that neutral gas is an important contributor to the overall gas mass found in the ISM of galaxies, including column density regions typical for molecular clouds.}
    
    \item {In order to further constrain the evolution of ${f(N_{\rm{H}_2})}$ additional observations and simulations are needed: At ${z=0}$ deeper observations are needed to constrain the low density end of ${f(N_{\rm{H}_2})}$. At ${z=3}$ high spatial resolution molecular gas observations of galaxies would enable the study of the high column density end of ${f(N_{\rm{H}_2})}$ and also probe more central regions of galaxies compared to absorption line studies. On the simulational side, efforts on resolving the cold gas phase within simulations are needed to constrain the high column density end of ${f(N_{\rm{H}_2})}$ at ${z=0}$. This may necessitate higher resolution simulations together with physical models for interstellar medium gas which aim to resolve the coldest phases. Further, the use of non-equilibrium chemical networks could provide a more accurate representation of the cold gas phase \citep[e.g.][]{Maio_2022}.}
    
\end{itemize}

\section*{Acknowledgements}
The Authors thank Jiayi Sun for the help with data from the PHANGS-ALMA survey and his comments and suggestions. We also would like to thank Sergei Balashev, Chia-Yu Hu, Gergö Popping and Thorsten Naab for the discussions, suggestions and comments that greatly helped to improve the paper. RS thanks ESO and the IMPRS program for the support of his PhD. DN acknowledges funding from the Deutsche Forschungsgemeinschaft (DFG) through an Emmy Noether Research Group (grant number NE 2441/1-1). The computations for the GRIFFIN project were carried out at CSC -- IT Center for Science Ltd. in Finland and the MPA cluster FREYA hosted by The Max Planck Computing and Data Facility (MPCDF) in Garching, Germany.

\section*{Data Availability}
Data directly related to this publication and its figures will be made available on request from the corresponding author. The IllustrisTNG simulations are publicly available and accessible in their entirety at \href{www.tng-project.org/data}{www.tng-project.org/data} \citep[][]{Nelson_2019b}.

\bibliographystyle{mnras}
\bibliography{example}

\begin{thebibliography}{}
\makeatletter
\relax
\def\mn@urlcharsother{\let\do\@makeother \do\$\do\&\do\#\do\^\do\_\do\%\do\~}
\def\mn@doi{\begingroup\mn@urlcharsother \@ifnextchar [ {\mn@doi@}
  {\mn@doi@[]}}
\def\mn@doi@[#1]#2{\def\@tempa{#1}\ifx\@tempa\@empty \href
  {http://dx.doi.org/#2} {doi:#2}\else \href {http://dx.doi.org/#2} {#1}\fi
  \endgroup}
\def\mn@eprint#1#2{\mn@eprint@#1:#2::\@nil}
\def\mn@eprint@arXiv#1{\href {http://arxiv.org/abs/#1} {{\tt arXiv:#1}}}
\def\mn@eprint@dblp#1{\href {http://dblp.uni-trier.de/rec/bibtex/#1.xml}
  {dblp:#1}}
\def\mn@eprint@#1:#2:#3:#4\@nil{\def\@tempa {#1}\def\@tempb {#2}\def\@tempc
  {#3}\ifx \@tempc \@empty \let \@tempc \@tempb \let \@tempb \@tempa \fi \ifx
  \@tempb \@empty \def\@tempb {arXiv}\fi \@ifundefined
  {mn@eprint@\@tempb}{\@tempb:\@tempc}{\expandafter \expandafter \csname
  mn@eprint@\@tempb\endcsname \expandafter{\@tempc}}}

\bibitem[\protect\citeauthoryear{{Altay}, {Theuns}, {Schaye}, {Crighton}  \&
  {Dalla Vecchia}}{{Altay} et~al.}{2011}]{Altay_2011}
{Altay} G.,  {Theuns} T.,  {Schaye} J.,  {Crighton} N. H.~M.,   {Dalla Vecchia}
  C.,  2011, \mn@doi [\apjl] {10.1088/2041-8205/737/2/L37}, \href
  {https://ui.adsabs.harvard.edu/abs/2011ApJ...737L..37A} {737, L37}

\bibitem[\protect\citeauthoryear{{Balashev} \& {Noterdaeme}}{{Balashev} \&
  {Noterdaeme}}{2018}]{Balashev_2018}
{Balashev} S.~A.,  {Noterdaeme} P.,  2018, \mn@doi [\mnras]
  {10.1093/mnrasl/sly067}, \href
  {https://ui.adsabs.harvard.edu/abs/2018MNRAS.478L...7B} {478, L7}

\bibitem[\protect\citeauthoryear{{Bigiel}, {Leroy}, {Walter}, {Brinks}, {de
  Blok}, {Madore}  \& {Thornley}}{{Bigiel} et~al.}{2008}]{Bigiel_2008}
{Bigiel} F.,  {Leroy} A.,  {Walter} F.,  {Brinks} E.,  {de Blok} W.~J.~G.,
  {Madore} B.,   {Thornley} M.~D.,  2008, \mn@doi [\aj]
  {10.1088/0004-6256/136/6/2846}, \href
  {https://ui.adsabs.harvard.edu/abs/2008AJ....136.2846B} {136, 2846}

\bibitem[\protect\citeauthoryear{{Bird}, {Vogelsberger}, {Haehnelt}, {Sijacki},
  {Genel}, {Torrey}, {Springel}  \& {Hernquist}}{{Bird}
  et~al.}{2014}]{Bird_2014}
{Bird} S.,  {Vogelsberger} M.,  {Haehnelt} M.,  {Sijacki} D.,  {Genel} S.,
  {Torrey} P.,  {Springel} V.,   {Hernquist} L.,  2014, \mn@doi [\mnras]
  {10.1093/mnras/stu1923}, \href
  {https://ui.adsabs.harvard.edu/abs/2014MNRAS.445.2313B} {445, 2313}

\bibitem[\protect\citeauthoryear{{Blitz} \& {Rosolowsky}}{{Blitz} \&
  {Rosolowsky}}{2006}]{Blitz_2006}
{Blitz} L.,  {Rosolowsky} E.,  2006, \mn@doi [\apj] {10.1086/505417}, \href
  {https://ui.adsabs.harvard.edu/abs/2006ApJ...650..933B} {650, 933}

\bibitem[\protect\citeauthoryear{{Bolatto}, {Wolfire}  \& {Leroy}}{{Bolatto}
  et~al.}{2013}]{Bolatto_2013}
{Bolatto} A.~D.,  {Wolfire} M.,   {Leroy} A.~K.,  2013, \mn@doi [\araa]
  {10.1146/annurev-astro-082812-140944}, \href
  {https://ui.adsabs.harvard.edu/abs/2013ARA&A..51..207B} {51, 207}

\bibitem[\protect\citeauthoryear{{Bonato} et~al.,}{{Bonato}
  et~al.}{2018}]{Bonato_2018}
{Bonato} M.,  et~al., 2018, \mn@doi [\mnras] {10.1093/mnras/sty1173}, \href
  {https://ui.adsabs.harvard.edu/abs/2018MNRAS.478.1512B} {478, 1512}

\bibitem[\protect\citeauthoryear{{Braun}}{{Braun}}{2012}]{Braun_2012}
{Braun} R.,  2012, \mn@doi [\apj] {10.1088/0004-637X/749/1/87}, \href
  {https://ui.adsabs.harvard.edu/abs/2012ApJ...749...87B} {749, 87}

\bibitem[\protect\citeauthoryear{{Christensen}, {M{\o}ller}, {Fynbo}  \&
  {Zafar}}{{Christensen} et~al.}{2014}]{Christensen_2014}
{Christensen} L.,  {M{\o}ller} P.,  {Fynbo} J.~P.~U.,   {Zafar} T.,  2014,
  \mn@doi [\mnras] {10.1093/mnras/stu1726}, \href
  {https://ui.adsabs.harvard.edu/abs/2014MNRAS.445..225C} {445, 225}

\bibitem[\protect\citeauthoryear{{Clark} \& {Glover}}{{Clark} \&
  {Glover}}{2014}]{Clark_2014}
{Clark} P.~C.,  {Glover} S. C.~O.,  2014, \mn@doi [\mnras]
  {10.1093/mnras/stu1589}, \href
  {https://ui.adsabs.harvard.edu/abs/2014MNRAS.444.2396C} {444, 2396}

\bibitem[\protect\citeauthoryear{{Cooke}, {Pettini}  \& {Steidel}}{{Cooke}
  et~al.}{2018}]{Cooke_2018}
{Cooke} R.~J.,  {Pettini} M.,   {Steidel} C.~C.,  2018, \mn@doi [\apj]
  {10.3847/1538-4357/aaab53}, \href
  {https://ui.adsabs.harvard.edu/abs/2018ApJ...855..102C} {855, 102}

\bibitem[\protect\citeauthoryear{{Crighton} et~al.,}{{Crighton}
  et~al.}{2015}]{Crighton_2015}
{Crighton} N. H.~M.,  et~al., 2015, \mn@doi [\mnras] {10.1093/mnras/stv1182},
  \href {https://ui.adsabs.harvard.edu/abs/2015MNRAS.452..217C} {452, 217}

\bibitem[\protect\citeauthoryear{{Decarli} et~al.,}{{Decarli}
  et~al.}{2020}]{Decarli_2020}
{Decarli} R.,  et~al., 2020, \mn@doi [\apj] {10.3847/1538-4357/abaa3b}, \href
  {https://ui.adsabs.harvard.edu/abs/2020ApJ...902..110D} {902, 110}

\bibitem[\protect\citeauthoryear{{Diemer} et~al.,}{{Diemer}
  et~al.}{2018}]{Diemer_2018}
{Diemer} B.,  et~al., 2018, \mn@doi [\apjs] {10.3847/1538-4365/aae387}, \href
  {https://ui.adsabs.harvard.edu/abs/2018ApJS..238...33D} {238, 33}

\bibitem[\protect\citeauthoryear{{Diemer} et~al.,}{{Diemer}
  et~al.}{2019}]{Diemer_2019}
{Diemer} B.,  et~al., 2019, \mn@doi [\mnras] {10.1093/mnras/stz1323}, \href
  {https://ui.adsabs.harvard.edu/abs/2019MNRAS.487.1529D} {487, 1529}

\bibitem[\protect\citeauthoryear{{Emsellem} et~al.,}{{Emsellem}
  et~al.}{2021}]{Emsellem_2021}
{Emsellem} E.,  et~al., 2021, arXiv e-prints, \href
  {https://ui.adsabs.harvard.edu/abs/2021arXiv211003708E} {p. arXiv:2110.03708}

\bibitem[\protect\citeauthoryear{{Feldmann}}{{Feldmann}}{2020}]{Feldman_2020}
{Feldmann} R.,  2020, \mn@doi [Communications Physics]
  {10.1038/s42005-020-00493-0}, \href
  {https://ui.adsabs.harvard.edu/abs/2020CmPhy...3..226F} {3, 226}

\bibitem[\protect\citeauthoryear{{French} et~al.,}{{French}
  et~al.}{2021}]{French_2021}
{French} D.~M.,  et~al., 2021, arXiv e-prints, \href
  {https://ui.adsabs.harvard.edu/abs/2021arXiv210807419F} {p. arXiv:2108.07419}

\bibitem[\protect\citeauthoryear{{Genel} et~al.,}{{Genel}
  et~al.}{2014}]{Genel_2014}
{Genel} S.,  et~al., 2014, \mn@doi [\mnras] {10.1093/mnras/stu1654}, \href
  {https://ui.adsabs.harvard.edu/abs/2014MNRAS.445..175G} {445, 175}

\bibitem[\protect\citeauthoryear{{Glover} \& {Clark}}{{Glover} \&
  {Clark}}{2012}]{Glover_2012}
{Glover} S. C.~O.,  {Clark} P.~C.,  2012, \mn@doi [\mnras]
  {10.1111/j.1365-2966.2011.20260.x}, \href
  {https://ui.adsabs.harvard.edu/abs/2012MNRAS.421..116G} {421, 116}

\bibitem[\protect\citeauthoryear{{Glover} \& {Mac Low}}{{Glover} \& {Mac
  Low}}{2007a}]{Glover_2007}
{Glover} S. C.~O.,  {Mac Low} M.-M.,  2007a, \mn@doi [\apjs] {10.1086/512238},
  \href {https://ui.adsabs.harvard.edu/abs/2007ApJS..169..239G} {169, 239}

\bibitem[\protect\citeauthoryear{{Glover} \& {Mac Low}}{{Glover} \& {Mac
  Low}}{2007b}]{Glover_2007b}
{Glover} S. C.~O.,  {Mac Low} M.-M.,  2007b, \mn@doi [\apj] {10.1086/512227},
  \href {https://ui.adsabs.harvard.edu/abs/2007ApJ...659.1317G} {659, 1317}

\bibitem[\protect\citeauthoryear{{Gnedin} \& {Kravtsov}}{{Gnedin} \&
  {Kravtsov}}{2011}]{Gnedin_2011}
{Gnedin} N.~Y.,  {Kravtsov} A.~V.,  2011, \mn@doi [\apj]
  {10.1088/0004-637X/728/2/88}, \href
  {https://ui.adsabs.harvard.edu/abs/2011ApJ...728...88G} {728, 88}

\bibitem[\protect\citeauthoryear{{Gnedin}, {Tassis}  \& {Kravtsov}}{{Gnedin}
  et~al.}{2009}]{Gnedin_2009}
{Gnedin} N.~Y.,  {Tassis} K.,   {Kravtsov} A.~V.,  2009, \mn@doi [\apj]
  {10.1088/0004-637X/697/1/55}, \href
  {https://ui.adsabs.harvard.edu/abs/2009ApJ...697...55G} {697, 55}

\bibitem[\protect\citeauthoryear{{Gong}, {Ostriker}  \& {Wolfire}}{{Gong}
  et~al.}{2017}]{Gong_2017}
{Gong} M.,  {Ostriker} E.~C.,   {Wolfire} M.~G.,  2017, \mn@doi [\apj]
  {10.3847/1538-4357/aa7561}, \href
  {https://ui.adsabs.harvard.edu/abs/2017ApJ...843...38G} {843, 38}

\bibitem[\protect\citeauthoryear{{Hamanowicz} et~al.,}{{Hamanowicz}
  et~al.}{2020}]{Hamanowicz_2020}
{Hamanowicz} A.,  et~al., 2020, \mn@doi [\mnras] {10.1093/mnras/stz3590}, \href
  {https://ui.adsabs.harvard.edu/abs/2020MNRAS.492.2347H} {492, 2347}

\bibitem[\protect\citeauthoryear{{Heintz}, {Watson}, {Oesch}, {Narayanan}  \&
  {Madden}}{{Heintz} et~al.}{2021}]{Heintz_2021}
{Heintz} K.~E.,  {Watson} D.,  {Oesch} P.~A.,  {Narayanan} D.,   {Madden}
  S.~C.,  2021, \mn@doi [\apj] {10.3847/1538-4357/ac2231}, \href
  {https://ui.adsabs.harvard.edu/abs/2021ApJ...922..147H} {922, 147}

\bibitem[\protect\citeauthoryear{{Hislop}, {Naab}, {Steinwandel}, {Lah{\'e}n},
  {Irodotou}, {Johansson}  \& {Walch}}{{Hislop} et~al.}{2021}]{Hislop_2021}
{Hislop} J.~M.,  {Naab} T.,  {Steinwandel} U.~P.,  {Lah{\'e}n} N.,  {Irodotou}
  D.,  {Johansson} P.~H.,   {Walch} S.,  2021, arXiv e-prints, \href
  {https://ui.adsabs.harvard.edu/abs/2021arXiv210908160H} {p. arXiv:2109.08160}

\bibitem[\protect\citeauthoryear{{Ho}, {Bird}  \& {Garnett}}{{Ho}
  et~al.}{2021}]{Ho_2021}
{Ho} M.-F.,  {Bird} S.,   {Garnett} R.,  2021, \mn@doi [\mnras]
  {10.1093/mnras/stab2169}, \href
  {https://ui.adsabs.harvard.edu/abs/2021MNRAS.507..704H} {507, 704}

\bibitem[\protect\citeauthoryear{{Hu}, {Naab}, {Walch}, {Moster}  \&
  {Oser}}{{Hu} et~al.}{2014}]{Hu_2014}
{Hu} C.-Y.,  {Naab} T.,  {Walch} S.,  {Moster} B.~P.,   {Oser} L.,  2014,
  \mn@doi [\mnras] {10.1093/mnras/stu1187}, \href
  {https://ui.adsabs.harvard.edu/abs/2014MNRAS.443.1173H} {443, 1173}

\bibitem[\protect\citeauthoryear{{Hu}, {Naab}, {Walch}, {Glover}  \&
  {Clark}}{{Hu} et~al.}{2016}]{Hu_2016}
{Hu} C.-Y.,  {Naab} T.,  {Walch} S.,  {Glover} S. C.~O.,   {Clark} P.~C.,
  2016, \mn@doi [\mnras] {10.1093/mnras/stw544}, \href
  {https://ui.adsabs.harvard.edu/abs/2016MNRAS.458.3528H} {458, 3528}

\bibitem[\protect\citeauthoryear{{Hu}, {Naab}, {Glover}, {Walch}  \&
  {Clark}}{{Hu} et~al.}{2017}]{Hu_2017}
{Hu} C.-Y.,  {Naab} T.,  {Glover} S. C.~O.,  {Walch} S.,   {Clark} P.~C.,
  2017, \mn@doi [\mnras] {10.1093/mnras/stx1773}, \href
  {https://ui.adsabs.harvard.edu/abs/2017MNRAS.471.2151H} {471, 2151}

\bibitem[\protect\citeauthoryear{{Hu}, {Sternberg}  \& {van Dishoeck}}{{Hu}
  et~al.}{2021}]{Hu_2021}
{Hu} C.-Y.,  {Sternberg} A.,   {van Dishoeck} E.~F.,  2021, arXiv e-prints,
  \href {https://ui.adsabs.harvard.edu/abs/2021arXiv210303889H} {p.
  arXiv:2103.03889}

\bibitem[\protect\citeauthoryear{{Jones}, {Haynes}, {Giovanelli}  \&
  {Moorman}}{{Jones} et~al.}{2018}]{Jones_2018}
{Jones} M.~G.,  {Haynes} M.~P.,  {Giovanelli} R.,   {Moorman} C.,  2018,
  \mn@doi [\mnras] {10.1093/mnras/sty521}, \href
  {https://ui.adsabs.harvard.edu/abs/2018MNRAS.477....2J} {477, 2}

\bibitem[\protect\citeauthoryear{{Klitsch}, {P{\'e}roux}, {Zwaan}, {Smail},
  {Oteo}, {Biggs}, {Popping}  \& {Swinbank}}{{Klitsch}
  et~al.}{2018}]{Klitsch_2018}
{Klitsch} A.,  {P{\'e}roux} C.,  {Zwaan} M.~A.,  {Smail} I.,  {Oteo} I.,
  {Biggs} A.~D.,  {Popping} G.,   {Swinbank} A.~M.,  2018, \mn@doi [\mnras]
  {10.1093/mnras/stx3184}, \href
  {https://ui.adsabs.harvard.edu/abs/2018MNRAS.475..492K} {475, 492}

\bibitem[\protect\citeauthoryear{{Klitsch} et~al.,}{{Klitsch}
  et~al.}{2019}]{Klitsch_2019}
{Klitsch} A.,  et~al., 2019, \mn@doi [\mnras] {10.1093/mnras/stz2660}, \href
  {https://ui.adsabs.harvard.edu/abs/2019MNRAS.490.1220K} {490, 1220}

\bibitem[\protect\citeauthoryear{{Krogager} \& {Noterdaeme}}{{Krogager} \&
  {Noterdaeme}}{2020}]{Krogager_2020}
{Krogager} J.-K.,  {Noterdaeme} P.,  2020, \mn@doi [\aap]
  {10.1051/0004-6361/202039843}, \href
  {https://ui.adsabs.harvard.edu/abs/2020A&A...644L...6K} {644, L6}

\bibitem[\protect\citeauthoryear{{Krogager}, {M{\o}ller}, {Fynbo}  \&
  {Noterdaeme}}{{Krogager} et~al.}{2017}]{Krogager_2017}
{Krogager} J.~K.,  {M{\o}ller} P.,  {Fynbo} J.~P.~U.,   {Noterdaeme} P.,  2017,
  \mn@doi [\mnras] {10.1093/mnras/stx1011}, \href
  {https://ui.adsabs.harvard.edu/abs/2017MNRAS.469.2959K} {469, 2959}

\bibitem[\protect\citeauthoryear{{Krumholz}}{{Krumholz}}{2013}]{Krumholz_2013}
{Krumholz} M.~R.,  2013, \mn@doi [\mnras] {10.1093/mnras/stt1780}, \href
  {https://ui.adsabs.harvard.edu/abs/2013MNRAS.436.2747K} {436, 2747}

\bibitem[\protect\citeauthoryear{{Lagos} et~al.,}{{Lagos}
  et~al.}{2015}]{Lagos_2015}
{Lagos} C. d.~P.,  et~al., 2015, \mn@doi [\mnras] {10.1093/mnras/stv1488},
  \href {https://ui.adsabs.harvard.edu/abs/2015MNRAS.452.3815L} {452, 3815}

\bibitem[\protect\citeauthoryear{{Lah{\'e}n}, {Naab}, {Johansson}, {Elmegreen},
  {Hu}  \& {Walch}}{{Lah{\'e}n} et~al.}{2019}]{Lahen_2019}
{Lah{\'e}n} N.,  {Naab} T.,  {Johansson} P.~H.,  {Elmegreen} B.,  {Hu} C.-Y.,
  {Walch} S.,  2019, \mn@doi [\apjl] {10.3847/2041-8213/ab2a13}, \href
  {https://ui.adsabs.harvard.edu/abs/2019ApJ...879L..18L} {879, L18}

\bibitem[\protect\citeauthoryear{{Lah{\'e}n}, {Naab}, {Johansson}, {Elmegreen},
  {Hu}, {Walch}, {Steinwandel}  \& {Moster}}{{Lah{\'e}n}
  et~al.}{2020a}]{Lahen_2020}
{Lah{\'e}n} N.,  {Naab} T.,  {Johansson} P.~H.,  {Elmegreen} B.,  {Hu} C.-Y.,
  {Walch} S.,  {Steinwandel} U.~P.,   {Moster} B.~P.,  2020a, \mn@doi [\apj]
  {10.3847/1538-4357/ab7190}, \href
  {https://ui.adsabs.harvard.edu/abs/2020ApJ...891....2L} {891, 2}

\bibitem[\protect\citeauthoryear{{Lah{\'e}n}, {Naab}, {Johansson}, {Elmegreen},
  {Hu}  \& {Walch}}{{Lah{\'e}n} et~al.}{2020b}]{Lahen_2020b}
{Lah{\'e}n} N.,  {Naab} T.,  {Johansson} P.~H.,  {Elmegreen} B.,  {Hu} C.-Y.,
  {Walch} S.,  2020b, \mn@doi [\apj] {10.3847/1538-4357/abc001}, \href
  {https://ui.adsabs.harvard.edu/abs/2020ApJ...904...71L} {904, 71}

\bibitem[\protect\citeauthoryear{{Lee} et~al.,}{{Lee} et~al.}{2021}]{Lee_2021}
{Lee} J.~C.,  et~al., 2021, arXiv e-prints, \href
  {https://ui.adsabs.harvard.edu/abs/2021arXiv210102855L} {p. arXiv:2101.02855}

\bibitem[\protect\citeauthoryear{{Leroy}, {Walter}, {Brinks}, {Bigiel}, {de
  Blok}, {Madore}  \& {Thornley}}{{Leroy} et~al.}{2008}]{Leroy_2008}
{Leroy} A.~K.,  {Walter} F.,  {Brinks} E.,  {Bigiel} F.,  {de Blok} W.~J.~G.,
  {Madore} B.,   {Thornley} M.~D.,  2008, \mn@doi [\aj]
  {10.1088/0004-6256/136/6/2782}, \href
  {https://ui.adsabs.harvard.edu/abs/2008AJ....136.2782L} {136, 2782}

\bibitem[\protect\citeauthoryear{{Leroy} et~al.,}{{Leroy}
  et~al.}{2013}]{Leroy_2013}
{Leroy} A.~K.,  et~al., 2013, \mn@doi [\aj] {10.1088/0004-6256/146/2/19}, \href
  {https://ui.adsabs.harvard.edu/abs/2013AJ....146...19L} {146, 19}

\bibitem[\protect\citeauthoryear{{Leroy} et~al.,}{{Leroy}
  et~al.}{2021}]{Leroy_2021}
{Leroy} A.~K.,  et~al., 2021, arXiv e-prints, \href
  {https://ui.adsabs.harvard.edu/abs/2021arXiv210407739L} {p. arXiv:2104.07739}

\bibitem[\protect\citeauthoryear{{Liu} et~al.,}{{Liu}
  et~al.}{2019}]{Daizhong_2019}
{Liu} D.,  et~al., 2019, \mn@doi [\apj] {10.3847/1538-4357/ab578d}, \href
  {https://ui.adsabs.harvard.edu/abs/2019ApJ...887..235L} {887, 235}

\bibitem[\protect\citeauthoryear{{Madau} \& {Dickinson}}{{Madau} \&
  {Dickinson}}{2014}]{Madau_2014}
{Madau} P.,  {Dickinson} M.,  2014, \mn@doi [\araa]
  {10.1146/annurev-astro-081811-125615}, \href
  {https://ui.adsabs.harvard.edu/abs/2014ARA&A..52..415M} {52, 415}

\bibitem[\protect\citeauthoryear{{Maio}, {P{\'e}roux}  \& {Ciardi}}{{Maio}
  et~al.}{2022}]{Maio_2022}
{Maio} U.,  {P{\'e}roux} C.,   {Ciardi} B.,  2022, \mn@doi [\aap]
  {10.1051/0004-6361/202142264}, \href
  {https://ui.adsabs.harvard.edu/abs/2022A&A...657A..47M} {657, A47}

\bibitem[\protect\citeauthoryear{{Marinacci} et~al.,}{{Marinacci}
  et~al.}{2018}]{Marinacci_2018}
{Marinacci} F.,  et~al., 2018, \mn@doi [\mnras] {10.1093/mnras/sty2206}, \href
  {https://ui.adsabs.harvard.edu/abs/2018MNRAS.480.5113M} {480, 5113}

\bibitem[\protect\citeauthoryear{{Mas-Ribas} et~al.,}{{Mas-Ribas}
  et~al.}{2017}]{Mas-Ribas_2017}
{Mas-Ribas} L.,  et~al., 2017, \mn@doi [\apj] {10.3847/1538-4357/aa81cf}, \href
  {https://ui.adsabs.harvard.edu/abs/2017ApJ...846....4M} {846, 4}

\bibitem[\protect\citeauthoryear{{Naiman} et~al.,}{{Naiman}
  et~al.}{2018}]{Naiman_2018}
{Naiman} J.~P.,  et~al., 2018, \mn@doi [\mnras] {10.1093/mnras/sty618}, \href
  {https://ui.adsabs.harvard.edu/abs/2018MNRAS.477.1206N} {477, 1206}

\bibitem[\protect\citeauthoryear{{Nelson} \& {Langer}}{{Nelson} \&
  {Langer}}{1997}]{Nelson_1997}
{Nelson} R.~P.,  {Langer} W.~D.,  1997, \mn@doi [\apj] {10.1086/304167}, \href
  {https://ui.adsabs.harvard.edu/abs/1997ApJ...482..796N} {482, 796}

\bibitem[\protect\citeauthoryear{{Nelson} et~al.,}{{Nelson}
  et~al.}{2018}]{Nelson_2018}
{Nelson} D.,  et~al., 2018, \mn@doi [\mnras] {10.1093/mnras/stx3040}, \href
  {https://ui.adsabs.harvard.edu/abs/2018MNRAS.475..624N} {475, 624}

\bibitem[\protect\citeauthoryear{{Nelson} et~al.,}{{Nelson}
  et~al.}{2019a}]{Nelson_2019b}
{Nelson} D.,  et~al., 2019a, \mn@doi [Computational Astrophysics and Cosmology]
  {10.1186/s40668-019-0028-x}, \href
  {https://ui.adsabs.harvard.edu/abs/2019ComAC...6....2N} {6, 2}

\bibitem[\protect\citeauthoryear{{Nelson} et~al.,}{{Nelson}
  et~al.}{2019b}]{Nelson_2019}
{Nelson} D.,  et~al., 2019b, \mn@doi [\mnras] {10.1093/mnras/stz2306}, \href
  {https://ui.adsabs.harvard.edu/abs/2019MNRAS.490.3234N} {490, 3234}

\bibitem[\protect\citeauthoryear{{Nilson}}{{Nilson}}{1973}]{Nilson_1973}
{Nilson} P.,  1973, {Uppsala general catalogue of galaxies}.
Uppsala Astron. Obs.

\bibitem[\protect\citeauthoryear{{Noterdaeme}, {Petitjean}, {Ledoux}  \&
  {Srianand}}{{Noterdaeme} et~al.}{2009}]{Noterdaeme_2009}
{Noterdaeme} P.,  {Petitjean} P.,  {Ledoux} C.,   {Srianand} R.,  2009, \mn@doi
  [\aap] {10.1051/0004-6361/200912768}, \href
  {https://ui.adsabs.harvard.edu/abs/2009A&A...505.1087N} {505, 1087}

\bibitem[\protect\citeauthoryear{{Noterdaeme} et~al.,}{{Noterdaeme}
  et~al.}{2012}]{Noterdaeme_2012}
{Noterdaeme} P.,  et~al., 2012, \mn@doi [\aap] {10.1051/0004-6361/201220259},
  \href {https://ui.adsabs.harvard.edu/abs/2012A&A...547L...1N} {547, L1}

\bibitem[\protect\citeauthoryear{{Oteo}, {Zwaan}, {Ivison}, {Smail}  \&
  {Biggs}}{{Oteo} et~al.}{2016}]{Otteo_2016}
{Oteo} I.,  {Zwaan} M.~A.,  {Ivison} R.~J.,  {Smail} I.,   {Biggs} A.~D.,
  2016, \mn@doi [\apj] {10.3847/0004-637X/822/1/36}, \href
  {https://ui.adsabs.harvard.edu/abs/2016ApJ...822...36O} {822, 36}

\bibitem[\protect\citeauthoryear{{P{\'e}roux} \& {Howk}}{{P{\'e}roux} \&
  {Howk}}{2020}]{Peroux_2020}
{P{\'e}roux} C.,  {Howk} J.~C.,  2020, \mn@doi [\araa]
  {10.1146/annurev-astro-021820-120014}, \href
  {https://ui.adsabs.harvard.edu/abs/2020ARA&A..58..363P} {58, 363}

\bibitem[\protect\citeauthoryear{{P{\'e}roux}, {Dessauges-Zavadsky},
  {D'Odorico}, {Sun Kim}  \& {McMahon}}{{P{\'e}roux}
  et~al.}{2005}]{Peroux_2005}
{P{\'e}roux} C.,  {Dessauges-Zavadsky} M.,  {D'Odorico} S.,  {Sun Kim} T.,
  {McMahon} R.~G.,  2005, \mn@doi [\mnras] {10.1111/j.1365-2966.2005.09432.x},
  \href {https://ui.adsabs.harvard.edu/abs/2005MNRAS.363..479P} {363, 479}

\bibitem[\protect\citeauthoryear{{P{\'e}roux}, {Bouch{\'e}}, {Kulkarni}, {York}
   \& {Vladilo}}{{P{\'e}roux} et~al.}{2011}]{Peroux_2011}
{P{\'e}roux} C.,  {Bouch{\'e}} N.,  {Kulkarni} V.~P.,  {York} D.~G.,
  {Vladilo} G.,  2011, \mn@doi [\mnras] {10.1111/j.1365-2966.2010.17598.x},
  \href {https://ui.adsabs.harvard.edu/abs/2011MNRAS.410.2237P} {410, 2237}

\bibitem[\protect\citeauthoryear{{Pessa} et~al.,}{{Pessa}
  et~al.}{2021}]{Pessa_2021}
{Pessa} I.,  et~al., 2021, \mn@doi [\aap] {10.1051/0004-6361/202140733}, \href
  {https://ui.adsabs.harvard.edu/abs/2021A&A...650A.134P} {650, A134}

\bibitem[\protect\citeauthoryear{{Pillepich} et~al.,}{{Pillepich}
  et~al.}{2018}]{Pillepich_2018}
{Pillepich} A.,  et~al., 2018, \mn@doi [\mnras] {10.1093/mnras/stx2656}, \href
  {https://ui.adsabs.harvard.edu/abs/2018MNRAS.473.4077P} {473, 4077}

\bibitem[\protect\citeauthoryear{{Pillepich} et~al.,}{{Pillepich}
  et~al.}{2019}]{Pillepich_2019}
{Pillepich} A.,  et~al., 2019, \mn@doi [\mnras] {10.1093/mnras/stz2338}, \href
  {https://ui.adsabs.harvard.edu/abs/2019MNRAS.490.3196P} {490, 3196}

\bibitem[\protect\citeauthoryear{{Planck Collaboration} et~al.,}{{Planck
  Collaboration} et~al.}{2016}]{Planck_2016}
{Planck Collaboration} et~al., 2016, \mn@doi [\aap]
  {10.1051/0004-6361/201525830}, \href
  {https://ui.adsabs.harvard.edu/abs/2016A&A...594A..13P} {594, A13}

\bibitem[\protect\citeauthoryear{{Popping} et~al.,}{{Popping}
  et~al.}{2019}]{Popping_2019}
{Popping} G.,  et~al., 2019, \mn@doi [\apj] {10.3847/1538-4357/ab30f2}, \href
  {https://ui.adsabs.harvard.edu/abs/2019ApJ...882..137P} {882, 137}

\bibitem[\protect\citeauthoryear{{Prochaska}, {O'Meara}  \&
  {Worseck}}{{Prochaska} et~al.}{2010}]{Prochaska_2010}
{Prochaska} J.~X.,  {O'Meara} J.~M.,   {Worseck} G.,  2010, \mn@doi [\apj]
  {10.1088/0004-637X/718/1/392}, \href
  {https://ui.adsabs.harvard.edu/abs/2010ApJ...718..392P} {718, 392}

\bibitem[\protect\citeauthoryear{{Rahmani} et~al.,}{{Rahmani}
  et~al.}{2018a}]{Rahmani_2018a}
{Rahmani} H.,  et~al., 2018a, \mn@doi [\mnras] {10.1093/mnras/stx2726}, \href
  {https://ui.adsabs.harvard.edu/abs/2018MNRAS.474..254R} {474, 254}

\bibitem[\protect\citeauthoryear{{Rahmani} et~al.,}{{Rahmani}
  et~al.}{2018b}]{Rahmani_2018b}
{Rahmani} H.,  et~al., 2018b, \mn@doi [\mnras] {10.1093/mnras/sty2216}, \href
  {https://ui.adsabs.harvard.edu/abs/2018MNRAS.480.5046R} {480, 5046}

\bibitem[\protect\citeauthoryear{{Rahmati}, {Pawlik}, {Rai{\v{c}}evi{\'c}}  \&
  {Schaye}}{{Rahmati} et~al.}{2013}]{Rahmati_2013}
{Rahmati} A.,  {Pawlik} A.~H.,  {Rai{\v{c}}evi{\'c}} M.,   {Schaye} J.,  2013,
  \mn@doi [\mnras] {10.1093/mnras/stt066}, \href
  {https://ui.adsabs.harvard.edu/abs/2013MNRAS.430.2427R} {430, 2427}

\bibitem[\protect\citeauthoryear{{Rathjen} et~al.,}{{Rathjen}
  et~al.}{2021}]{Rathjen_2021}
{Rathjen} T.-E.,  et~al., 2021, \mn@doi [\mnras] {10.1093/mnras/stab900}, \href
  {https://ui.adsabs.harvard.edu/abs/2021MNRAS.504.1039R} {504, 1039}

\bibitem[\protect\citeauthoryear{{Richings} \& {Schaye}}{{Richings} \&
  {Schaye}}{2016}]{Richings_2016}
{Richings} A.~J.,  {Schaye} J.,  2016, \mn@doi [\mnras] {10.1093/mnras/stw327},
  \href {https://ui.adsabs.harvard.edu/abs/2016MNRAS.458..270R} {458, 270}

\bibitem[\protect\citeauthoryear{{Riechers} et~al.,}{{Riechers}
  et~al.}{2019}]{Riechers_2019}
{Riechers} D.~A.,  et~al., 2019, \mn@doi [\apj] {10.3847/1538-4357/aafc27},
  \href {https://ui.adsabs.harvard.edu/abs/2019ApJ...872....7R} {872, 7}

\bibitem[\protect\citeauthoryear{{R{\"o}ttgers}, {Naab}, {Cernetic},
  {Dav{\'e}}, {Kauffmann}, {Borthakur}  \& {Foidl}}{{R{\"o}ttgers}
  et~al.}{2020}]{Roettgers_2020}
{R{\"o}ttgers} B.,  {Naab} T.,  {Cernetic} M.,  {Dav{\'e}} R.,  {Kauffmann} G.,
   {Borthakur} S.,   {Foidl} H.,  2020, \mn@doi [\mnras]
  {10.1093/mnras/staa1490}, \href
  {https://ui.adsabs.harvard.edu/abs/2020MNRAS.496..152R} {496, 152}

\bibitem[\protect\citeauthoryear{{Saintonge} \& {Catinella}}{{Saintonge} \&
  {Catinella}}{2022}]{Saintonge_2022}
{Saintonge} A.,  {Catinella} B.,  2022, arXiv e-prints, \href
  {https://ui.adsabs.harvard.edu/abs/2022arXiv220200690S} {p. arXiv:2202.00690}

\bibitem[\protect\citeauthoryear{{Saintonge} et~al.,}{{Saintonge}
  et~al.}{2011}]{Saintonge_2011}
{Saintonge} A.,  et~al., 2011, \mn@doi [\mnras]
  {10.1111/j.1365-2966.2011.18677.x}, \href
  {https://ui.adsabs.harvard.edu/abs/2011MNRAS.415...32S} {415, 32}

\bibitem[\protect\citeauthoryear{{S{\'a}nchez} et~al.,}{{S{\'a}nchez}
  et~al.}{2014}]{Sanchez_2014}
{S{\'a}nchez} S.~F.,  et~al., 2014, \mn@doi [\aap]
  {10.1051/0004-6361/201322343}, \href
  {https://ui.adsabs.harvard.edu/abs/2014A&A...563A..49S} {563, A49}

\bibitem[\protect\citeauthoryear{{S{\'a}nchez} et~al.,}{{S{\'a}nchez}
  et~al.}{2019}]{Sanchez_2019}
{S{\'a}nchez} S.~F.,  et~al., 2019, \mn@doi [\mnras] {10.1093/mnras/stz019},
  \href {https://ui.adsabs.harvard.edu/abs/2019MNRAS.484.3042S} {484, 3042}

\bibitem[\protect\citeauthoryear{{Schaye}}{{Schaye}}{2001}]{Schaye_2001}
{Schaye} J.,  2001, \mn@doi [\apjl] {10.1086/338106}, \href
  {https://ui.adsabs.harvard.edu/abs/2001ApJ...562L..95S} {562, L95}

\bibitem[\protect\citeauthoryear{{Schaye} et~al.,}{{Schaye}
  et~al.}{2015}]{Schaye_2015}
{Schaye} J.,  et~al., 2015, \mn@doi [\mnras] {10.1093/mnras/stu2058}, \href
  {https://ui.adsabs.harvard.edu/abs/2015MNRAS.446..521S} {446, 521}

\bibitem[\protect\citeauthoryear{{Schroetter} et~al.,}{{Schroetter}
  et~al.}{2019}]{Schroetter_2019}
{Schroetter} I.,  et~al., 2019, \mn@doi [\mnras] {10.1093/mnras/stz2822}, \href
  {https://ui.adsabs.harvard.edu/abs/2019MNRAS.490.4368S} {490, 4368}

\bibitem[\protect\citeauthoryear{{Seifried} et~al.,}{{Seifried}
  et~al.}{2017}]{Seifried_2017}
{Seifried} D.,  et~al., 2017, \mn@doi [\mnras] {10.1093/mnras/stx2343}, \href
  {https://ui.adsabs.harvard.edu/abs/2017MNRAS.472.4797S} {472, 4797}

\bibitem[\protect\citeauthoryear{{Sijacki}, {Vogelsberger}, {Genel},
  {Springel}, {Torrey}, {Snyder}, {Nelson}  \& {Hernquist}}{{Sijacki}
  et~al.}{2015}]{Sijacki_2015}
{Sijacki} D.,  {Vogelsberger} M.,  {Genel} S.,  {Springel} V.,  {Torrey} P.,
  {Snyder} G.~F.,  {Nelson} D.,   {Hernquist} L.,  2015, \mn@doi [\mnras]
  {10.1093/mnras/stv1340}, \href
  {https://ui.adsabs.harvard.edu/abs/2015MNRAS.452..575S} {452, 575}

\bibitem[\protect\citeauthoryear{{Spilker}, {Kainulainen}  \&
  {Orkisz}}{{Spilker} et~al.}{2021}]{Spilker_2021}
{Spilker} A.,  {Kainulainen} J.,   {Orkisz} J.,  2021, arXiv e-prints, \href
  {https://ui.adsabs.harvard.edu/abs/2021arXiv210804518S} {p. arXiv:2108.04518}

\bibitem[\protect\citeauthoryear{{Springel}}{{Springel}}{2005}]{Springel_2005}
{Springel} V.,  2005, \mn@doi [\mnras] {10.1111/j.1365-2966.2005.09655.x},
  \href {https://ui.adsabs.harvard.edu/abs/2005MNRAS.364.1105S} {364, 1105}

\bibitem[\protect\citeauthoryear{{Springel}}{{Springel}}{2010}]{Springel_2010}
{Springel} V.,  2010, \mn@doi [\mnras] {10.1111/j.1365-2966.2009.15715.x},
  \href {https://ui.adsabs.harvard.edu/abs/2010MNRAS.401..791S} {401, 791}

\bibitem[\protect\citeauthoryear{{Springel} \& {Hernquist}}{{Springel} \&
  {Hernquist}}{2003}]{Springel_2003}
{Springel} V.,  {Hernquist} L.,  2003, \mn@doi [\mnras]
  {10.1046/j.1365-8711.2003.06206.x}, \href
  {https://ui.adsabs.harvard.edu/abs/2003MNRAS.339..289S} {339, 289}

\bibitem[\protect\citeauthoryear{{Springel} et~al.,}{{Springel}
  et~al.}{2018}]{Springel_2018}
{Springel} V.,  et~al., 2018, \mn@doi [\mnras] {10.1093/mnras/stx3304}, \href
  {https://ui.adsabs.harvard.edu/abs/2018MNRAS.475..676S} {475, 676}

\bibitem[\protect\citeauthoryear{Stevens, Diemer, Lagos, Nelson, Obreschkow,
  Wang  \& Marinacci}{Stevens et~al.}{2019}]{Stevens_2019}
Stevens A. R.~H.,  Diemer B.,  Lagos C. d.~P.,  Nelson D.,  Obreschkow D.,
  Wang J.,   Marinacci F.,  2019, \mn@doi [Monthly Notices of the Royal
  Astronomical Society] {10.1093/mnras/stz2513}, 490, 96–113

\bibitem[\protect\citeauthoryear{{Sun} et~al.,}{{Sun} et~al.}{2018}]{Sun_2018}
{Sun} J.,  et~al., 2018, \mn@doi [\apj] {10.3847/1538-4357/aac326}, \href
  {https://ui.adsabs.harvard.edu/abs/2018ApJ...860..172S} {860, 172}

\bibitem[\protect\citeauthoryear{{Sun} et~al.,}{{Sun} et~al.}{2020}]{Sun_2020}
{Sun} J.,  et~al., 2020, \mn@doi [\apjl] {10.3847/2041-8213/abb3be}, \href
  {https://ui.adsabs.harvard.edu/abs/2020ApJ...901L...8S} {901, L8}

\bibitem[\protect\citeauthoryear{{Szakacs} et~al.,}{{Szakacs}
  et~al.}{2021}]{Szakacs_2021}
{Szakacs} R.,  et~al., 2021, \mn@doi [\mnras] {10.1093/mnras/stab1434}, \href
  {https://ui.adsabs.harvard.edu/abs/2021MNRAS.505.4746S} {505, 4746}

\bibitem[\protect\citeauthoryear{{Tacconi} et~al.,}{{Tacconi}
  et~al.}{2018}]{Tacconi_2018}
{Tacconi} L.~J.,  et~al., 2018, \mn@doi [\apj] {10.3847/1538-4357/aaa4b4},
  \href {https://ui.adsabs.harvard.edu/abs/2018ApJ...853..179T} {853, 179}

\bibitem[\protect\citeauthoryear{{Tacconi}, {Genzel}  \& {Sternberg}}{{Tacconi}
  et~al.}{2020}]{Tacconi_2020}
{Tacconi} L.~J.,  {Genzel} R.,   {Sternberg} A.,  2020, \mn@doi [\araa]
  {10.1146/annurev-astro-082812-141034}, \href
  {https://ui.adsabs.harvard.edu/abs/2020ARA&A..58..157T} {58, 157}

\bibitem[\protect\citeauthoryear{{Villaescusa-Navarro}
  et~al.,}{{Villaescusa-Navarro} et~al.}{2018}]{Villaescusa-Navarro_2018}
{Villaescusa-Navarro} F.,  et~al., 2018, \mn@doi [\apj]
  {10.3847/1538-4357/aadba0}, \href
  {https://ui.adsabs.harvard.edu/abs/2018ApJ...866..135V} {866, 135}

\bibitem[\protect\citeauthoryear{{Vogelsberger} et~al.,}{{Vogelsberger}
  et~al.}{2014a}]{Vogelsberger_2014}
{Vogelsberger} M.,  et~al., 2014a, \mn@doi [\mnras] {10.1093/mnras/stu1536},
  \href {https://ui.adsabs.harvard.edu/abs/2014MNRAS.444.1518V} {444, 1518}

\bibitem[\protect\citeauthoryear{{Vogelsberger} et~al.,}{{Vogelsberger}
  et~al.}{2014b}]{Vogelsberger_2014b}
{Vogelsberger} M.,  et~al., 2014b, \mn@doi [\nat] {10.1038/nature13316}, \href
  {https://ui.adsabs.harvard.edu/abs/2014Natur.509..177V} {509, 177}

\bibitem[\protect\citeauthoryear{{Walch} et~al.,}{{Walch}
  et~al.}{2015}]{Walch_2015}
{Walch} S.,  et~al., 2015, \mn@doi [\mnras] {10.1093/mnras/stv1975}, \href
  {https://ui.adsabs.harvard.edu/abs/2015MNRAS.454..238W} {454, 238}

\bibitem[\protect\citeauthoryear{{Walter} et~al.,}{{Walter}
  et~al.}{2020}]{Walter_2020}
{Walter} F.,  et~al., 2020, \mn@doi [\apj] {10.3847/1538-4357/abb82e}, \href
  {https://ui.adsabs.harvard.edu/abs/2020ApJ...902..111W} {902, 111}

\bibitem[\protect\citeauthoryear{{Weigel}, {Schawinski}  \&
  {Bruderer}}{{Weigel} et~al.}{2016}]{Weigel_2016}
{Weigel} A.~K.,  {Schawinski} K.,   {Bruderer} C.,  2016, \mn@doi [\mnras]
  {10.1093/mnras/stw756}, \href
  {https://ui.adsabs.harvard.edu/abs/2016MNRAS.459.2150W} {459, 2150}

\bibitem[\protect\citeauthoryear{{Weinberger} et~al.,}{{Weinberger}
  et~al.}{2017}]{Weinberger_2017}
{Weinberger} R.,  et~al., 2017, \mn@doi [\mnras] {10.1093/mnras/stw2944}, \href
  {https://ui.adsabs.harvard.edu/abs/2017MNRAS.465.3291W} {465, 3291}

\bibitem[\protect\citeauthoryear{{Wolfe}, {Gawiser}  \& {Prochaska}}{{Wolfe}
  et~al.}{2005}]{Wolfe_2005}
{Wolfe} A.~M.,  {Gawiser} E.,   {Prochaska} J.~X.,  2005, \mn@doi [\araa]
  {10.1146/annurev.astro.42.053102.133950}, \href
  {https://ui.adsabs.harvard.edu/abs/2005ARA&A..43..861W} {43, 861}

\bibitem[\protect\citeauthoryear{{Yates}, {P{\'e}roux}  \& {Nelson}}{{Yates}
  et~al.}{2021}]{Yates_2021}
{Yates} R.~M.,  {P{\'e}roux} C.,   {Nelson} D.,  2021, \mn@doi [\mnras]
  {10.1093/mnras/stab2837}, \href
  {https://ui.adsabs.harvard.edu/abs/2021MNRAS.508.3535Y} {508, 3535}

\bibitem[\protect\citeauthoryear{{Zabl} et~al.,}{{Zabl}
  et~al.}{2020}]{Zabl_2020}
{Zabl} J.,  et~al., 2020, \mn@doi [\mnras] {10.1093/mnras/stz3607}, \href
  {https://ui.adsabs.harvard.edu/abs/2020MNRAS.492.4576Z} {492, 4576}

\bibitem[\protect\citeauthoryear{{Zafar}, {P{\'e}roux}, {Popping}, {Milliard},
  {Deharveng}  \& {Frank}}{{Zafar} et~al.}{2013}]{Zafar_2013}
{Zafar} T.,  {P{\'e}roux} C.,  {Popping} A.,  {Milliard} B.,  {Deharveng}
  J.~M.,   {Frank} S.,  2013, \mn@doi [\aap] {10.1051/0004-6361/201321154},
  \href {https://ui.adsabs.harvard.edu/abs/2013A&A...556A.141Z} {556, A141}

\bibitem[\protect\citeauthoryear{{Zwaan}}{{Zwaan}}{2000}]{Zwaan_2000}
{Zwaan} M.~A.,  2000, PhD thesis, -

\bibitem[\protect\citeauthoryear{{Zwaan} \& {Prochaska}}{{Zwaan} \&
  {Prochaska}}{2006}]{Zwaan_2006}
{Zwaan} M.~A.,  {Prochaska} J.~X.,  2006, \mn@doi [\apj] {10.1086/503191},
  \href {https://ui.adsabs.harvard.edu/abs/2006ApJ...643..675Z} {643, 675}

\bibitem[\protect\citeauthoryear{{Zwaan}, {van der Hulst}, {Briggs},
  {Verheijen}  \& {Ryan-Weber}}{{Zwaan} et~al.}{2005}]{Zwaan_2005}
{Zwaan} M.~A.,  {van der Hulst} J.~M.,  {Briggs} F.~H.,  {Verheijen} M.~A.~W.,
   {Ryan-Weber} E.~V.,  2005, \mn@doi [\mnras]
  {10.1111/j.1365-2966.2005.09698.x}, \href
  {https://ui.adsabs.harvard.edu/abs/2005MNRAS.364.1467Z} {364, 1467}

\bibitem[\protect\citeauthoryear{{den Brok} et~al.,}{{den Brok}
  et~al.}{2021}]{den_Brok_2021}
{den Brok} J.~S.,  et~al., 2021, \mn@doi [\mnras] {10.1093/mnras/stab859},
  \href {https://ui.adsabs.harvard.edu/abs/2021MNRAS.504.3221D} {504, 3221}

\bibitem[\protect\citeauthoryear{{van de Voort}, {Springel}, {Mandelker}, {van
  den Bosch}  \& {Pakmor}}{{van de Voort} et~al.}{2019}]{vandevoort_2019}
{van de Voort} F.,  {Springel} V.,  {Mandelker} N.,  {van den Bosch} F.~C.,
  {Pakmor} R.,  2019, \mn@doi [\mnras] {10.1093/mnrasl/sly190}, \href
  {https://ui.adsabs.harvard.edu/abs/2019MNRAS.482L..85V} {482, L85}

\bibitem[\protect\citeauthoryear{{van der Hulst}, {van Albada}  \&
  {Sancisi}}{{van der Hulst} et~al.}{2001}]{van_der_Hulst_2001}
{van der Hulst} J.~M.,  {van Albada} T.~S.,   {Sancisi} R.,  2001, in {Hibbard}
  J.~E.,  {Rupen} M.,   {van Gorkom} J.~H.,  eds,  Astronomical Society of the
  Pacific Conference Series Vol. 240, Gas and Galaxy Evolution. p.~451

\makeatother
\end{thebibliography}

\appendix

\section{\texorpdfstring{$\bm{f(N_{\rm{H}_2})}$}{f(N\_H2)} Dependence on Physical Properties}

\label{app:cdd_dep}

We explore how the integrated properties of galaxies in the PHANGS-ALMA sample shape the $f(N_{\rm{H}_2}$) of individual objects. Namely we study the dependence of the $f(N_{\rm{H}_2}$) on the star formation rate (SFR), stellar mass ($M_*$) and H$_2$ mass ($M_{\rm{H}_2}$). 

The colour coding in Fig. \ref{fig:f_N_PHANGS_individ} already displays the dependence of the $f(N_{\rm{H}_2}$) on the three parameters mentioned above. In order to quantify this relationship, we fit a gamma distribution of the form:

\begin{equation}
     f(N_{\rm{H}_2}) = \frac{f^*}{N^*} \;  \left(\frac{N_{\rm{H}_2}}{N^*}\right)^{-\beta} \; \rm{e}^{-\frac{N_{\rm{H}_2}}{N^*}} \; \; ,
\end{equation}

\noindent to the computed individual $f(N_{\rm{H}_2}$) of the 150 pc resolution PHANGS-ALMA sample. Note that there is no physical motivation for fitting a gamma distribution to the individual $f(N_{\rm{H}_2}$), it simply provides good fits of the $f(N_{\rm{H}_2}$) for a minimal number of parameters.

\begin{figure*}
  \includegraphics[width=1.0\textwidth]{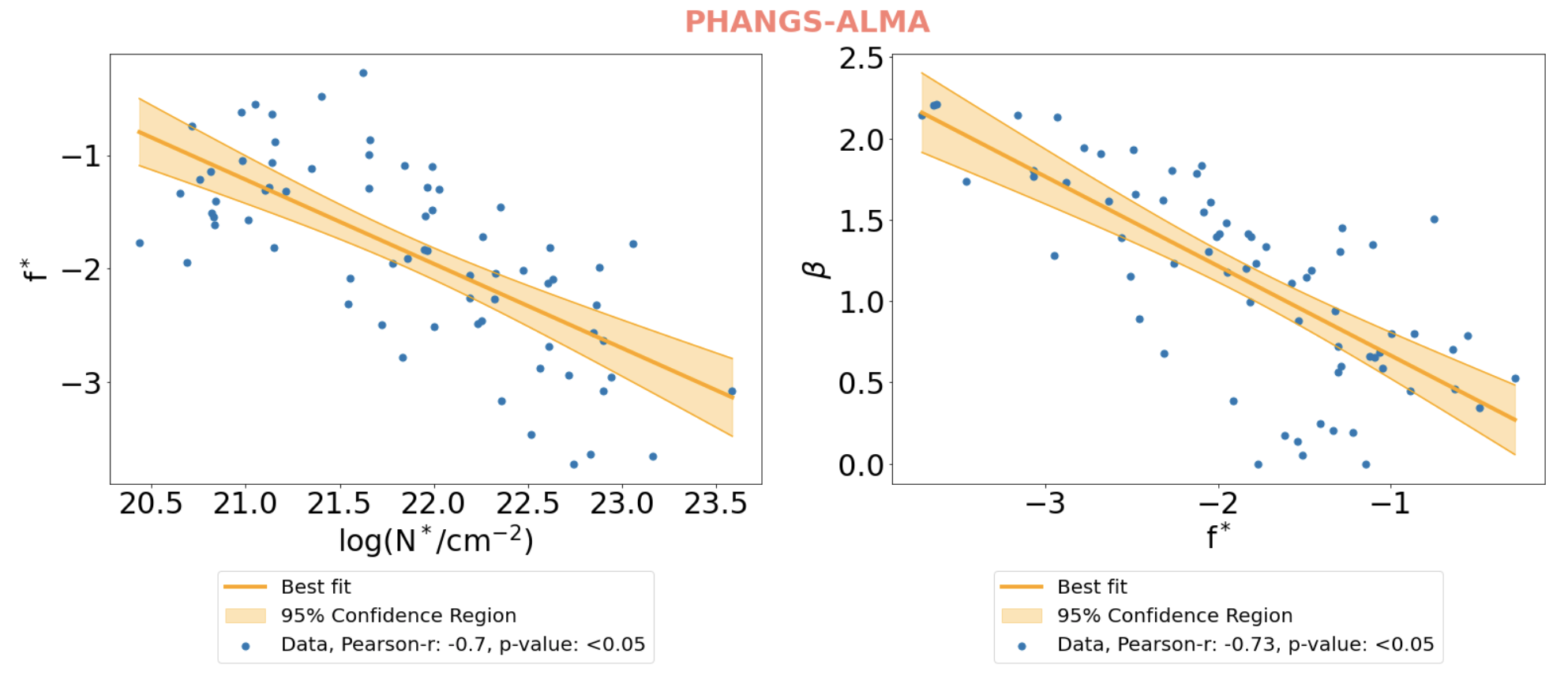}
  \caption{Correlations between the free parameter $f^*$ and the free parameter $N^*$ and between $f^*$ and the slope $\beta$ of the gamma distribution fits on the $f(N_{\rm{H}_2}$) in the PHANGS-ALMA sample. The blue dots indicate the fit values of individual galaxies, the orange line the best fit and the orange band the 95\% confidence region of the fit. Both samples show strong correlations with Pearson-rs of $\sim 0.7$ and p-values < 0.05.} 
  \label{fig:f_N_PHANGS_individ_N_f_dep}
\end{figure*}

The individual $f(N_{\rm{H}_2}$) is largely determined by the parameter $N^*$ as the second free parameter $f^*$ correlates with $N^*$ (slope: $-0.74 \pm 0.09$,
intercept: $14.3\pm2.0$, Pearson-r: 0.7, p-value [calculated using a Kolmogorov-Smirnov test] <0.05) and $\beta$ in turn correlates with $f^*$ (slope: $-0.55 \pm 0.06$,
intercept $0.11 \pm 0.13$, Pearson-r:0.73, p-value <0.05). Fig. \ref{fig:f_N_PHANGS_individ_N_f_dep} displays the dependence of $f^*$ on $N^*$ and $\beta$ on $f^*$. 

\begin{figure*}
  \includegraphics[width=1.0\textwidth]{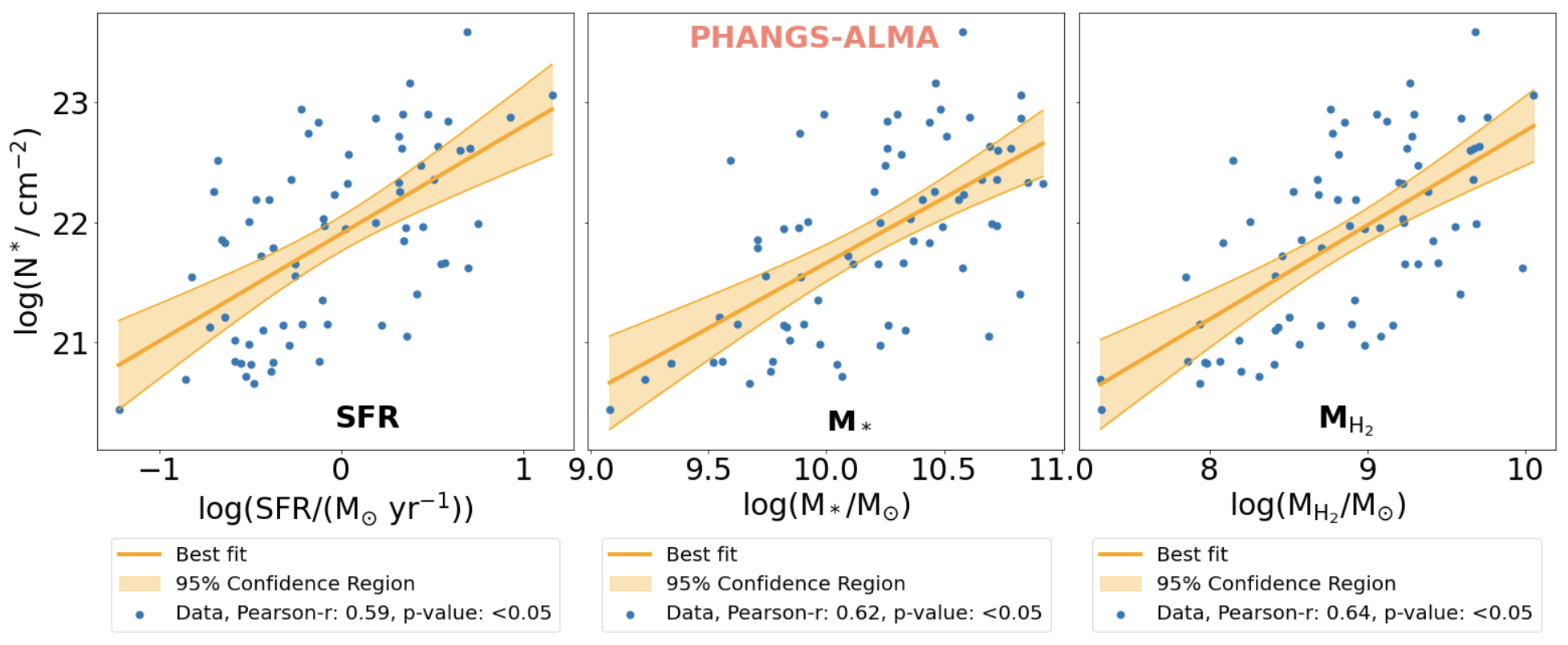}
  \caption{Dependence on different physical properties of the free parameter $N^*$ of the gamma distribution fits on the $f(N_{\rm{H}_2}$) within the PHANGS-ALMA sample. The blue dots indicate the fit values and properties of individual objects, the orange line the best fit and the orange band the 95\% confidence region of the fit. The physical properties (SFR, $M_*$ and $M_{\rm{H_2}}$) of the galaxies correlate with the free parameter $N^*$, albeit with a significant scatter.} 
  \label{fig:f_N_PHANGS_individ_N_dep_x}
\end{figure*}

As already indicated in Fig. \ref{fig:f_N_PHANGS_individ}, $f(N_{\rm{H}_2}$) depends on physical parameters of the galaxies within the PHANGS-ALMA sample. This is quantified in Fig. \ref{fig:f_N_PHANGS_individ_N_dep_x}, where we show the relationship of the free parameter $N^*$ of the gamma distribution with SFR,  $M_*$ and $\Sigma_{\rm{H_2}}$. As $N^*$ largely determines the $f(N_{\rm{H}_2}$) of a galaxy, it is implied that these physical properties of a galaxy affect the $f(N_{\rm{H}_2}$) of a galaxy. The three studied properties of the galaxies show the following correlation and fit parameters (in log space): SFR - $N^*$: slope: $0.90 \pm 0.15$ intercept: $21.90 \pm 0.07$, Pearson-r = 0.59, p-value < 0.05; $M_*$ - $N^*$: slope: $1.21 \pm 0.05$ intercept: $-5.5 \pm 1.1$ Pearson-r = 0.62, p-value < 0.05 and $M_{\rm{H_2}}$ - $N^*$: slope: $0.79 \pm 0.11$, intercept: $14.9 \pm 1.0$ Pearson-r: 0.64, p-value < 0.05.

Using the SFR, $M_*$ or $M_{\rm{H_2}}$ of a galaxy one could approximate its $f(N_{\rm{H}_2}$) using these correlations. We note that our tests have shown that while these fits approximate the global $f(N_{\rm{H}_2}$) well when using the PHANGS-ALMA sample, they often fail for individual galaxies because they are degenerate. We therefore caution from using these fits to predict individual $f(N_{\rm{H}_2}$).

\bsp
\label{lastpage}
\end{document}